\documentclass[journal]{vgtc}                     

\usepackage{enumitem}  
\usepackage{graphicx}  
\usepackage{algorithm}
\usepackage{amsmath}
\usepackage{amsfonts}
\usepackage{algpseudocode} 

\onlineid{1494}

\vgtccategory{Data Transformations}

\vgtcpapertype{}

\title{Localized Evaluation for Constructing Discrete Vector Fields}

\author{%
  Tanner Finken, 
  Julien Tierny, and
  Joshua A. Levine
}

\authorfooter{
  \item
  	Tanner Finken is with University of Arizona. 
  	E-mail: finkent@arizona.edu.
   \item 
        Julien Tierny is with Sorbonne University. \\
        E-mail: julien.tierny@sorbonne-universite.fr.
    \item 
        Joshua A.~Levine is with University of Arizona. \\
        E-mail: josh@cs.arizona.edu.
   
}

\abstract{
Topological abstractions offer a method to summarize the behavior of vector fields, but computing them robustly can be challenging due to numerical precision issues.
One alternative is to represent the vector field using a discrete approach, which constructs a collection of pairs of simplices in the input mesh that satisfies criteria introduced by Forman's discrete Morse theory.
While numerous approaches exist to compute pairs in the restricted case of the gradient of a scalar field, state-of-the-art algorithms for the general case of vector fields require expensive optimization procedures.
This paper introduces a fast, novel approach for pairing simplices of two-dimensional, triangulated vector fields that do not vary in time.
The key insight of our approach is that we can employ a local evaluation, inspired by the approach used to construct a discrete gradient field, where every
\jaledit{simplex} 
in a mesh is considered by no more than one of its vertices.
Specifically, we observe that for any edge in the input mesh, we can uniquely assign an outward direction of flow.
We can further expand this consistent notion of outward flow at each vertex, which corresponds to the concept of a downhill flow in the case of scalar fields.
Working with outward flow enables a linear-time algorithm that processes the (outward) neighborhoods of each vertex one-by-one, similar to the approach used for scalar fields.
We couple our approach to constructing discrete vector fields with a method to extract, simplify, and visualize topological features.
Empirical results on analytic and simulation data demonstrate drastic improvements in running time, produce features similar to the current state-of-the-art, and show the application of simplification to large, complex flows.

}


\newcommand{\hidefigure}[1]{FIGURE HERE}

\teaser{
  \centering
  \includegraphics[alt={A map of earth with flow along oceans and points in various locations.},width=0.98\linewidth]
  {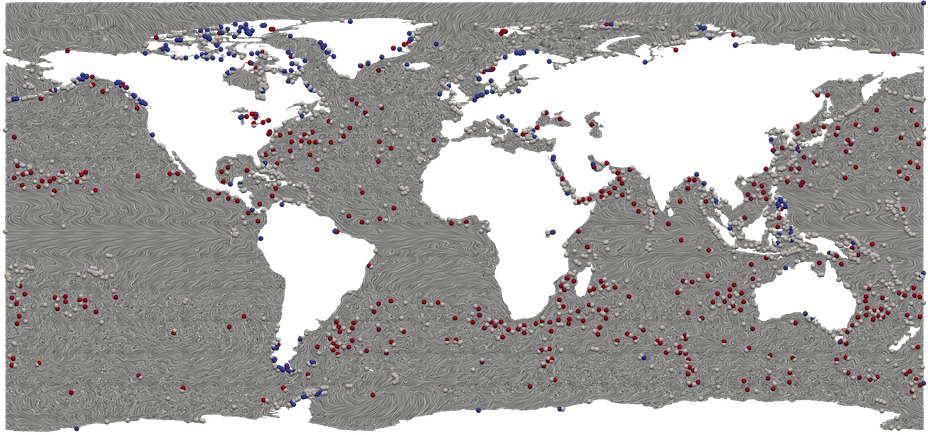}
  \caption{
   We extract and simplify a vector field of ocean currents using our technique. 
 The input mesh has >48 million simplices, and the original flow results in over 65000 critical points.  We simplify to approximately 2000 critical points using a discrete representation of 
 \change{the} field.  Computing the field for a domain this big takes only 4 minutes and simplification takes approximately 10 minutes.
  \label{fig:teaser}
  }
}

\graphicspath{{figs/}{figures/}{pictures/}{images/}{./}} 

\usepackage{tabu}                      
\usepackage{booktabs}                  
\usepackage{lipsum}                    
\usepackage{mwe}                       
\usepackage{subcaption}
\usepackage{mathptmx}                  
\newcommand{\St}{\mathrm{St}}
\newcommand{\outStar}{\mathrm{St}^\leftrightarrow}
\newcommand{\R}{\mathbb{R}}
\newcommand{\change}[1]{#1}
\newcommand{\jaledit}[1]{#1}
\newcommand{\highlight}[1]{\textcolor{blue!80}{#1}}

\begin{document}


\firstsection{Introduction}

\maketitle

Vector fields are a fundamental way to encode dynamic systems from a variety of applications, including aeronautics, climate, energy, geoscience, and materials science.  
Visualization of vector fields can help to understand the intricate patterns inherent in these systems, and such insights can be used to improve designs and validate computational simulations relative to physical observations.
Due to the complex ways in which the behavior of flowing phenomena can manifest, a desirable approach is to \change{visualize} abstractions from topological data analysis that can summarize flow patterns.
Unfortunately, computing such abstractions directly from an input vector field can be difficult when the field is represented as samples on an input grid or mesh, a common output from numerical simulation. 
An appealing solution is to construct discrete representations, which trade expensive numerical operations (such as integrating a streamline) for robust, discrete counterparts.  

Notably, Forman's discrete Morse theory~\cite{Forman01} introduces the notion of a \emph{discrete vector field}, wherein the vector field is represented by a collection of pairs of simplices in an input mesh.  
Pairs encode a notion of flow along the field on discrete intervals, specifically at the granularity of the individual mesh elements.  
Pairs are also required to follow specific rules (such as the dimension of the simplices involved in each pair must differ by one), and under such rules one can reason about the properties of the field itself.   
This representation enables the powerful concepts of Morse theory to be translated to practical, computable data structures.
The main challenge for constructing a discrete vector field becomes choosing an appropriate collection of pairs based on the given input vector field.
\jaledit{While numerous algorithms can efficiently (in $O(n)$) construct a discrete gradient field, the more general case of discrete vector fields, in which rotation and cycles can manifest, remains only partially understood.} 
Particularly, state-of-the-art approaches, such as the FastCVT method of Reininghaus et al.~\cite{fastCVT}, still require $O(n^{\frac{3}{2}}\log n)$ computation due to a global optimization involving repeated computations of shortest paths.

Nevertheless, if one can produce a discrete vector field, the downstream payoff could be significant.  
Topological abstractions become relatively easy to compute, \change{e.g.,}~critical points in a discrete vector field correspond to unmatched simplices and orbits in a discrete vector field correspond to paths (in discrete Morse theory, referred to as \emph{V-paths}) which traverse the same pair more than once.  
One does not need to rely on numerical integration methods or fiddle with numerical tolerances during computation.  
Moreover, discrete representations naturally enable algorithms to rank and simplify such topological features, enabling a multiscale analysis of the flow.
The popularity of discrete representations in topological data analysis for scalar fields is growing for similar reasons, and the research community is continuing to demonstrate their utility and practicality~\cite{gyulassy_vis08,LowerStar,guillou_tvcg23}.  
In part, this work is inspired by the success of discrete Morse theory applied to scalar fields, and our investigation took motivation from  both their popularity and the specific algorithmic choices used for constructing discrete gradient fields.



The goal of this work is to efficiently compute a discrete vector field (alternatively, a collection of pairs of simplices)\footnote{N.B., we prefer the terminology of discrete vector field 
\jaledit{instead of} \emph{combinatorial vector field}~\cite{forman1998combinatorial}.  The two concepts are 
\jaledit{similar} since a discrete vector field is described by pairing simplices, while a combinatorial vector field is more readily viewed through the lens of graph theory with the set of pairs corresponding to a matching on the simplicial graph.  A key practical difference appears to be if V-paths may flow through the entire simplicial graph (as alluded to in Section 6 of Forman~\cite{Forman01} and optimized for in Reininghaus et al.~\cite{exactCVT,fastCVT}) or if they are restricted to, as is more typical, subcomplexes corresponding to specific pairs of dimensions.  
\jaledit{Our} preference for ``discrete'' comes from desiring intuition consistent with topological data analysis of scalar fields.}, and we specifically focus on the case of two-dimensional, triangulated vector fields.
We develop an approach for the case of \emph{steady} vector fields, that is vector fields which do not vary in time.
While there are numerous possible ways to construct a collection of pairs, we would also like to ensure that the discrete representation captures a geometric notion of similarity between the input field and its discrete representation.   
Looking over to how one constructs a discrete gradient field, the most efficient algorithms use an approach based on a localized evaluation of similarity, evaluating the neighborhood of each vertex to determine which portion of it has a lower function value (more formally: the \emph{lower star} of each vertex).
Lower stars are sufficient to partition the simplices of an input mesh, and thus one can make a consistent evaluation for assigning pairs by looking only at individual neighborhoods.

Since a general vector field can have rotation, there would appear to be no global equivalent.
Specifically, one cannot choose a scalar field for which the concept of ``lower'' can be made to align with an arbitrary vector field (i.e., one cannot just ``integrate'' the vector field to produce a scalar field whose gradient matches the vector field).
Nevertheless, our key insight is based on the observation that it is possible to choose a best fit alignment if one restricts themselves to a local evaluation of flow structure.
Ultimately, this allows our algorithm to make a consistent choice on which simplices are flowing \emph{outward} from each vertex.  
In this sense, outward flow corresponds to a localized definition of the concept of lower or downhill.
While outward stars will not produce a complete partition of the simplices, they will result in a consistent one.  
Said another way: no simplex will be assigned to the outward star of more than one vertex, but some simplices may not belong into any outward star.
Perhaps unsurprisingly, simplices that are left out correspond to criticalities in the vector field, and these will naturally be left as unpaired since they will be skipped in assignment to vertices.

We can then use outward flow as an ingredient to apply the simpler, more efficient algorithms employed for computing discrete gradient fields that also capture geometric closeness.  
Specifically, with minor modifications we can employ homotopy expansion~\cite{LowerStar} on the outward stars of vertices, resulting in a fast and geometrically accurate computation of a discrete vector field. 
Using these discrete vector fields, one can then extract topological summaries.  
Interestingly, the localized concept of outward naturally leads to a relative notion of magnitude of change along an arbitrary $V$-path in the discrete flow.
This means that given a specific $V$-path between two criticalities, we can accumulate a (relative) change similar to the concept of \emph{persistence} used to rank pairs of critical points and simplify scalar field topology~\cite{edelsbrunner02}.
While this is only appropriate for certain pairs of simplices, we can use this to simplify the topology of the discrete vector field as well.
Specifically, saddle-source, saddle-sink, and saddle-orbit pairs can both be ranked and simplified in this manner.

\subsection*{Contributions}
We summarize our contributions as
\begin{enumerate}
    \item We develop a new algorithm for computing discrete vector fields, inspired by the homotopy expansion approach used when computing scalar field topology and based on a localized evaluation of outward.
    \item Using the resulting discrete vector fields, we extract,  simplify, and visualize topological abstractions.
    \item We compare our approach to existing techniques for discrete and piecewise linear vector fields, both in terms of speed and accuracy.
\end{enumerate}

Our experimental work is backed by an implementation of this new method, as well as the method of Reininghaus et al.~\cite{fastCVT} using the Topology ToolKit (TTK)~\cite{TiernyFLGM18}, so as to compare both methods fairly.  Ultimately, we anticipate adding this code to a future release of TTK.




\section{Related Work}


Our work builds on recent advancements in both vector field visualization and scalar field topology.  We review both fields.

\subsection{Vector Field Visualization with Topological Summaries}

Topological summaries have been a key tool for visualizing the structure of vector fields.  Helman and Hesselink introduced the use of a topological skeleton and their use in visualization in the early 1990s\cite{HelmanH89,HelmanH90}.  In the case of a steady vector field, this structure captures the structure of streamlines by providing a decomposition of the domain in terms of \emph{separatrices}, which 
differentiate bundles of streamlines in terms of their origin and destination.  Together with extracting  closed orbits~\cite{wischgoll:closedorbits} and classifying critical points~\cite{perry1987description}, this approach can summarize the flow behavior.  These structures have also been used in \change{a} variety of more complex cases, including 
flow in higher dimensions\cite{GlobusLL91} and 
time-varying flow\cite{TheiselWHS04,UffingerSE13}.  Multiple recent articles and surveys cover the use of topological structures~\cite{GarthT05,LHZP07,PobitzerPFSKTMH11,ScheuermannCH04,wang2016numerics} for flow visualization, and we refer the reader to them for a complete coverage of this topic.

In this work, we focus explicitly on the representation choice one can make for encoding flow, and its downstream impact on extracting topological structures.  Multiple authors have relied on interesting discretizations of flow due to their increased robustness, and in this work we make a similar tradeoff.

One line of work, due to Chen et al., constructs a graph that encodes where the image of individual triangles flow to when advected by flow~\cite{chen2007vector,ChenMLZ08}.  Using these graphs, computed from an input piecewise linear vector field, one can construct the Morse decomposition, which captures a structural representation similar to topological skeletons.  Through the application of Conley index theorem, the result\change{ing} Morse sets can then be classified~\cite{ChenDSLZ12}.  A similar approach can \change{be} used for input piecewise constant vector fields~\cite{szymczak2011stable}.  Like our work, the graph constructed represents a discrete representation of flow, but in our case we make choices at the level of individual simplices, following the approach of discrete Morse theory.

A second approach is to explicitly model the flow across edges in an input mesh, using the concept of an edge map~\cite{BJBCLNP12}. Edge maps decompose edges to explicitly track where flow travels across individual triangles, and one can structure a graph that tracks which subset of an edge flows where. For piecewise linear flow, the combinations can be quite complex~\cite{JBBLNP11}, although in certain settings can bound the errors between this combinatorial representation and the underlying flow.  Alternatively, instead of discretizing edges at the level of subsets, one can explicitly quantize each edge uniformly~\cite{LJBPB12}, which creates much larger graphs but also can improve accuracy as well as enable the extraction of topological features.  This approach models streamlines explicitly, rather than collapsing flow to the discrete concept of V-paths, but also requires both large graphs and computations.

Finally, a combinatorial vector field approach has been proposed as a direct translation of discrete Morse theory~\cite{exactCVT}.  This approach constructs a complete simplicial graph, representing adjacency through faces of an input simplicial complex.  A matching on this graph is then optimized to reflect a global geometric criterion, which inspired our use of a similar local measure.  As this is expensive to compute, Reininghaus et al. next proposed an improved optimization method that is also amenable to parallelization~\cite{fastCVT}.  In practice, this method is quite similar to ours, and we directly compare it to our work to demonstrate improved running times by avoiding a global optimization.  They also employ a method for simplification, based on a global ordering of cancellations through their optimization procedure.

While not explicitly focused on discrete representations, we also mention the work of Skraba et al.~which introduced the concept of \emph{robustness} for critical points for 2D~\cite{SkrabaWCR14} and 3D~\cite{SkrabaRWCBP16} steady vector fields.  Based on the concept of a well diagram~\cite{chazal2012computing}, this work can provide tools for ranking and simplifying critical points, similar in spirit to how persistence in used for scalar field simplification.  While multiple works have targeted topological simplification of vector fields~\cite{de1999collapsing,de1999visualization,tricoche2000topology,tricoche2001continuous}, there have been few works that offer a complete toolset which provides visualization and analysis tools comparable to those available for scalar fields.



\subsection{Scalar Field Topology}

Separately, topological tools have met with great success in analyzing scalar field data.  Our work takes inspiration from some of the computational methods used in the simpler domain of scalar fields. In particular, we rely on the tools from discrete Morse theory~\cite{forman1998combinatorial} for representing flow, and, as such, we would like to leverage the local nature of how discrete gradient fields are constructed. 

We review some of the key applications of topological data analysis for scalar fields.  
Topological methods for scalar fields have been extensively studied by the visualization community over the last two decades \cite{heine16}. 
One of their key advantages is their ability to map physical phenomena from a variety \change{of} domains into a common framework of features.  For example, authors have used topological segmentation to model dissipation elements in combustion applications~\cite{gyulassy_ev14}, vortices in fluid dynamics~\cite{kasten_tvcg11, nauleau_ldav22}, and skeletal structures in medical images~\cite{topoAngler}.  
Topological connectivity can be used to model the connections in the cosmic web~\cite{sousbie11, shivashankar2016felix}, atomic level connections in chemistry~\cite{harshChemistry, Malgorzata19, olejniczak_pccp23}, and functional connectivity from MRI~\cite{beiBrain18}. 
Topological simplification is a shared component in many applications, and has been noted as key enabling for scaling up to large scale simulations~\cite{bremer_tvcg11, gyulassy_vis15}.


We focus our discussion on the use of the Morse-Smale complex as a topological abstraction for representing scalar fields.  
The Morse-Smale complex partitions the domain of a scalar field relative to gradient flow.  
Early algorithms for computing the Morse-Smale complex utilized numerical primitives to capture gradient flow~\cite{edelsbrunner03}.  
While implementing these algorithms in practice was achievable in 2D~\cite{timo03,timo04}, extending the 3D algorithm~\cite{edelsbrunner03b} proved more difficult. 
A key to making these algorithms practical was employing discrete Morse theory~\cite{forman1998combinatorial} to represent the input using a discrete gradient field~\cite{gyulassy_vis08}.  
As research progressed~\cite{LowerStar, ShivashankarN12}, Gyulassy et al.~described a generic kernel for processing neighborhoods~\cite{gyulassy_vis12}, which observed a key property previously shown by Robins et al.~\cite{LowerStar}.  
Specifically, that if one takes a greedy approach to assignment, this can impact the overall geometry as represented by the discrete flow, and moreover if one fails to satisfy a homotopy expansion this can create spurious critical points.
We thus adopt a similar property, processing vertex neighborhoods in an order similar to Robins et al.~\cite{LowerStar}.

Researchers have continued to pioneer the use of Morse-Smale complexes computed with discrete representations, improving on speed and accuracy~\cite{Defl15, gyulassy_vis18, robin23}.  
Thanks to several importance metrics~\cite{edelsbrunner02, carr04}, these abstractions can also be iteratively simplified, hence enabling multi-scale feature analysis.
A key to simplification is the use of persistent homology, as captured by the persistence diagram~\cite{edelsbrunner02, dipha} that ranks features of different dimensions and associates them with critical points~\cite{banchoff70}.
Even recent approaches for computing persistence diagrams have relied on the benefits of a discrete gradient field, such as the discrete Morse sandwich approach of Guillou et al.~\cite{guillou_tvcg23}. 
By framing the problem of vector field topology using a similar metaphor, we also hope to \jaledit{enable}
a multi-scale simplification method.

\section{Background}

In this section, we define our notation regarding a vector field defined on a simplicial complex and discuss related concepts in the piecewise linear case. 
Next, we switch to foundation work in discrete Morse theory and provide background on both the mathematical definitions and practicalities of how one computes them.
In the case of a discrete vector field, we provide \change{numerical relations on} how the piecewise linear\change{, continuous} flow is captured by discrete flow.
Finally, we describe key aspects of how discrete gradient fields are computed, connecting these concepts to \change{our relation on discrete flow to apply for }the more general case of discrete vector fields.

\subsection{Vector Fields on Simplicial Complexes}

In general, a \emph{$d$-dimensional (steady) vector field} is defined by a function $F\colon \mathcal{M} \rightarrow \R^d$, such that each point $x \in \mathcal{M}$ has a $d$-dimensional vector $F(x)$.
\change{For} piecewise linear vector fields, we assume that the domain $\mathcal{M}$ is a $d'$-dimensional simplicial complex $K$\change{, with $F$ represented by vectors stored at}
the vertices of $K$.  
$F(x)$ for all other points $x$ on the interior of simplices can then be computed through linear interpolation of the vectors on the corners.
In general, $d = d'$ is not required, but in this work\change{,} we will assume that $d = d' = 2$.
Thus, we focus on the two-dimensional setting, and $K$ corresponds to a triangulation of $\mathcal{M}$ and vectors on the interiors of triangles are the result of interpolating the three vectors on the corners.

Topological features of interest for vector fields can be described through the concept of a streamline.
Formally, a \emph{streamline} through position $x_0$ is the solution to the equation $\frac{\partial x}{\partial t} = F(x)$ with the initial condition $x(0) = x_0$.
Streamlines 
\change{represent} how massless particles travel through $\mathcal{M}$ if their instantaneous velocity vector matches $F$.
\change{If} $F(x) = 0$, the streamline will reduce \change{to} a point, and we call such points \emph{critical points}.
Otherwise, in the generic case as $t$ approaches both $\infty$ and $-\infty$, a streamline will trace a 1-dimensional path whose limit points are either critical points or exit the boundary.
It is also possible for a streamline to form a \emph{closed orbit}, in which case the path loops back on itself (\change{i.e.,} if there exist $t_1$ and $t_2$ such that $x(t_1) = x(t_2)$, which will produce a path that repeats indefinitely).

Critical points for vector fields can be classified by the flow patterns in a small neighborhood around them~\cite{perry1987description}.
In two-dimensions, the complete classification captures whether they are attracting/stable (referred to as a \emph{sink}), repelling/unstable (a \emph{source}), or neither (a \emph{saddle}).  
As vector fields may exhibit curl, these critical points may also 
\change{contain} rotation, resulting in rotating versions of sources (an attracting \emph{focus}) and sinks (a repelling \emph{focus}).  
Finally, the neighborhood around a critical point may be purely rotating (a \emph{center}).
In the neighborhood around a saddle, most streamlines will avoid the critical point, but there are four canonical streamlines, referred to as \emph{separatrices}, which provide asymptotic boundaries where the flow switches.
Together with orbits, the separatrices of saddles form the \emph{topological skeleton}~\cite{HelmanH89,HelmanH90}.

Finally, we review terminology for simplicial complexes.  
A \emph{$p$-dimensional simplex} is a convex combination of $p+1$ points embedded within the domain $\mathcal{M}$.
We abuse notation slightly and will refer to simplices in both the abstract sense (\change{e.g.,} a set of points) as well as their geometric realization.
Where relevant, we denote the dimension $p$ of a simplex $\sigma$ as $\sigma^{(p)}$. 
For $\sigma^{(p)} = \{v_0, v_1, \ldots, v_p\}$, a face of a $\sigma^{(p)}$ is a subset of its vertices, which by definition is also a simplex (in the case of proper subsets, a simplex of a reduced dimension).
We write $\tau \leq \sigma$ to indicate that $\tau$ is a face of $\sigma$, and likewise say $\sigma$ is a \emph{coface} of $\tau$.
A \emph{simplicial complex}, $K$, is then a set of simplices such that (a) for any $\sigma \in K$, all faces of $\sigma$ are also in $K$ and (b) for any $\sigma, \tau \in K$, if $\sigma \cap \tau \neq \emptyset$ then $\sigma \cap \tau$ is a face of both $\sigma$ and $\tau$.
Given a simplicial complex $K$, we can define a notion of a neighborhood for a simplex.  
Specifically, given a simplex $\sigma$, the \emph{star} of $\sigma$ is its set of cofaces in $K$, notated as  $\St(\sigma) = \{ \alpha \ | \ \sigma \leq \alpha \}$.

\subsection{Discrete Morse Theory}

Discrete Morse theory provides a translation of concepts of Morse theory in the smooth case to discrete objects like simplicial complexes~\cite{Forman01}.  
Given a simplicial complex $K$, a \emph{discrete vector} is a pair of simplices $(\alpha^{(p)}, \beta^{(p+1)})$ whose dimension differs by 1 and for which $\alpha < \beta$. 
Pairs capture directionality; we think of the pair as going from $\alpha$ (the ``tail'') to $\beta$ (the ``head'').
A \emph{discrete vector field}, $V$, on $K$ is a collection of discrete vectors such that each simplex is involved in no more than one pair.  

Many of the concepts from the continuous case of vector fields can be translated to this discrete framework.  
Any simplex that is unpaired is a \emph{critical simplex}, which correspond to the concept of critical points in the vector field.
The dimensionality of a critical simplex $\sigma^{(p)}$ reflects certain aspects of its classification, and we refer to it as the \emph{index} of the critical simplex.  
Specifically, index 0 corresponds to attracting, index 1 correspond to saddle-like, and index 2 corresponds to repelling behavior.
Note that this type description is complete for scalar fields, but as vector fields include rotation it is insufficient to capture all possible behaviors.

Likewise, the notion of a streamline is replaced with the concept following sequences of simplices in the direction indicated by the pairs.
Specifically, a \emph{$V$-path}, $P$, of index $p$ is a sequence of simplices $$\alpha^{(p)}_0, \beta^{(p+1)}_0, \alpha^{(p)}_1, \beta^{(p+1)}_1, \ldots$$ such that $(\alpha^{(p)}_i, \beta^{(p+1)}_i) \in V$ and $\alpha^{(p)}_{i+1}$ is a face of $\beta^{(p+1)}_i$ not selected as a pair (for convenience, we refer to $\beta^{(p+1)}_i$ and $\alpha^{(p)}_{i+1}$ as an \emph{anti-pair}).

Separatrices can be modeled as $V$-paths for which we include the critical (unpaired) simplices as the start and end nodes of $V$-path, to reflect the behavior of flow in the limit.
%
%
In discrete vector fields, it \change{is possible for} $V$-paths \change{to contain repeated pairs.}
In this case, we say the $V$-path is a \emph{closed orbit} of index $p$.
A discrete vector field that does not exhibit orbits is a \emph{discrete gradient field}, matching the notion that gradient fields are irrotational.
Following Reininghaus et al.\cite{fastCVT}, if the index of a closed orbit is 0, we say it is an \emph{attracting orbit}, and if the index is 1, we call it a \emph{repelling orbit}.
Similarly, separatrices can also connect saddles to orbits.

\subsection{\change{Relating Continuous Flow with Discrete Flow} }

Using a discrete vector field presents numerous computational advantages.  
Notably, one can replace the concept of computing streamlines (which require numeric integration techniques to solve the differential equation) with a combinatorial operation of tracing a streamline.
Detecting critical points is a simple lookup, rather than requiring a root finding operation to identify their exact location.
That said, given a piecewise linear vector field, a question of interest in this work is how to best define a discrete vector field on the same simplicial complex while simultaneously capturing the input field as close as possible, subject to granularity at which the simplicial complex covers the domain.

To answer this question, 
\jaledit{we consider} 
the approach used for constructing a discrete gradient field from an input scalar field. 
Most algorithms~\cite{gyulassy_vis12,LowerStar} apply a variant of processing vertex neighbors, based on utilizing the star of each vertex. 
Given a scalar field defined by the function 
\jaledit{$g()$, a lower star $L(x)$ for a vertex $x$ is defined as:}
\begin{equation}
    L(x) = \{ \sigma \in \St(x) \ | \ g(x) = \max_{v_i \in \sigma} \  g(v_i) \}.
\end{equation} 
Intuitively, the lower star of a vertex encompasses all 
\jaledit{simplices} 
in its star that have the vertex as the highest value, implying a 
\jaledit{decrease} 
in scalar value when moving \jaledit{from $x$ to} 
any 
\jaledit{simplex} 
$\sigma \in L(x)$.

Robins et al. take the approach of processing the lower star of each vertex to assign pairs in a unique order that maintains the topological structure~\cite{LowerStar}. 
Note that, since the lower stars of vertices form a partition of $K$, each vertex can assign pairs completely independently.
Thus, each vertex can be processed independently and the entire algorithm is embarrassingly parallel.
Furthermore, for scalar fields, one can show that each piecewise linear critical point (which will always lie on input vertices) will manifest as a nearby discrete critical 
\jaledit{simplex} 
in its star~\cite{LowerStar,TiernyFLGM18}.

To devise a similar algorithm for vector fields, one needs to replicate certain key concepts on which this algorithm is based. 
We leverage the definitions of \emph{weight} used by Reininghaus et al.~\cite{fastCVT}.
This definition of weight captures a coarse-grained value analogous to integrating the vector field over a short distance (along a simplex).
Thus, weight intuitively relates to the concept of change in magnitude, allowing us to determine, in a localized way, which simplices are ``lower''. 

\jaledit{We introduce this definition by first defining how we interpret the vector field for simplices.  For a simplex $\sigma$, let $c(\sigma)$ be the barycenter (or average of the coordinates of its vertices) of $\sigma$.  We evaluate $F$ on $c(\sigma)$ when considering weight.  Assuming piecewise linear interpolation, this means $F(c(\sigma))$ will be the average of the vectors on the vertex coordinates.}



The weight of a given link (possible pair) from $\alpha ^{(p)}$ to $\beta ^{(p+1)}$ measures the alignment of a discrete vector with $F$.
Specifically, we compare the geometric representations of the simplices involved in the discrete pair to their corresponding values of $F$.
Weight captures the averaged values of $F$ for the simplices involved with the pair, dotted with a vector associated with moving from \change{the barycenter of }$\change{\alpha}$ to \change{the barycenter of} $\change{\beta})$.  
\begin{equation}
    \label{eq:linkWeight}
    w(\alpha, \beta) = \frac{\jaledit{F(c(\alpha))+F(c(\beta))}}{2} \cdot (c(\beta) - c(\alpha)).
\end{equation} 

Finally, this definition can be extended to determine the weight of a $V$-path by summing the \emph{alternating} set of weights along the sequence, summing terms $w(\alpha_i^{(p)}, \beta_i^{(p+1)})$ and $-w(\beta_i^{(p+1)}, \alpha_{i+1}^{(p)})$.
This sign of terms alternates to reflect that the direction of moving along a $V$-path aligns with the geometry 
when following a pair and reverses from the geometry when following an anti-pair.  
Given that separatrices may also terminate at critical points and/or orbits, this means that the weight of a $V$-path, $P$, can be written as:
\begin{equation}
    \label{eq:vpath}
    w(P) = \sum w(\alpha_i^{(p)}, \beta_i^{(p+1)}) - \sum w(\beta_i^{(p+1)}, \alpha_{i+1}^{(p)}).
\end{equation}
Note that given a $V$-path corresponding to an orbit, we stop at the \change{final anti-pair before the repeated pair because that pair will not be reversed during simplification as described later. } 

\section{Our Approach}

We define a concept of outward flow so as to categorize the star of each vertex in the input simplicial complex for the vector field.  Then from this definition we are able to process each vertex through using a modified version of the Robins et al.'s algorithm ProcessLowerStars~\cite{LowerStar}. 

\subsection{Defining Outward Star}
\begin{figure}[!t]
    \centering
    \begin{subfigure}[b]{0.31\linewidth}
      \center{\includegraphics[alt={An image of a line segment with vectors coming out each side and the center. The center vector points up and slightly right indicating flow from left to right on the line. },width=\linewidth]{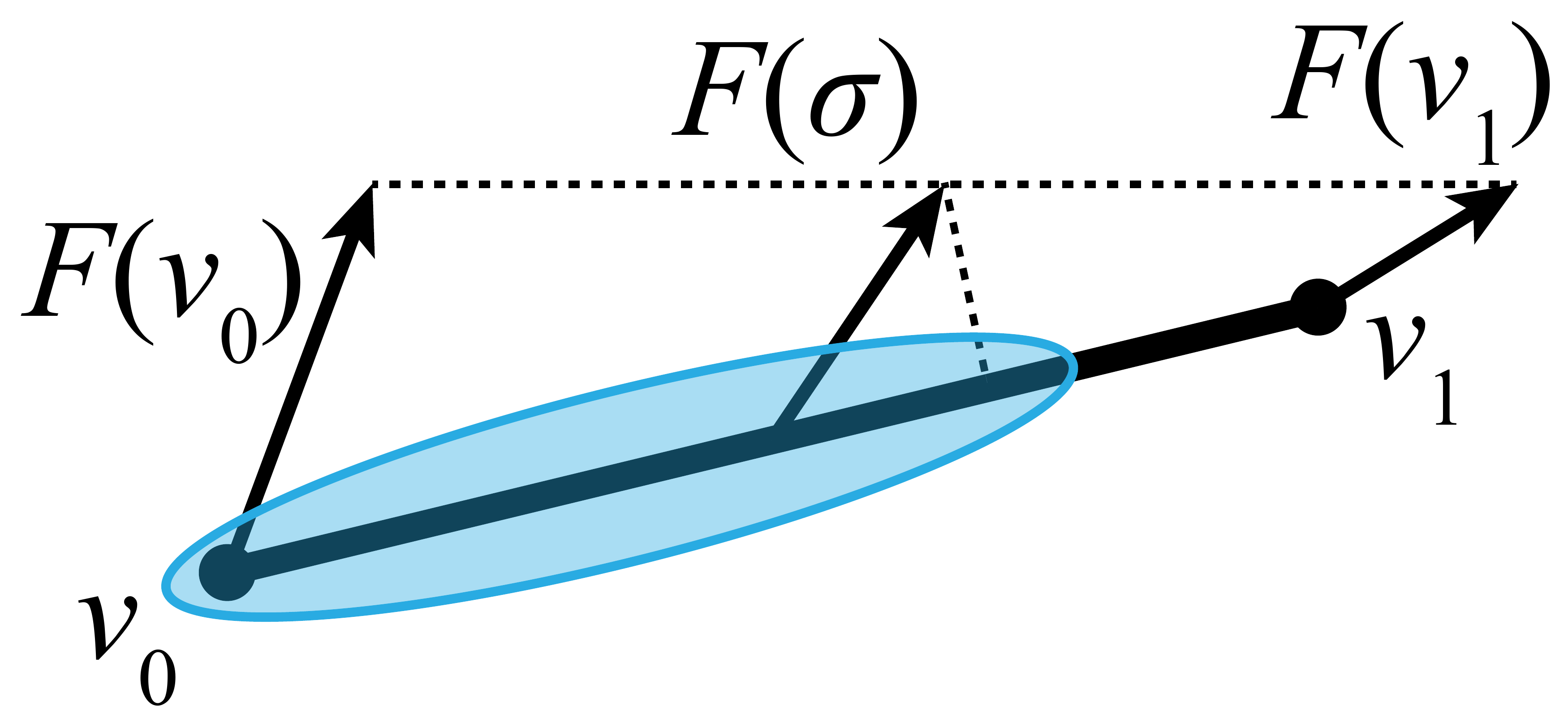}}
      \vspace{-1em}
      \caption{}
      \vspace{-1em}
    \end{subfigure}
    \begin{subfigure}[b]{0.31\linewidth}
      \center{\includegraphics[width=\linewidth]{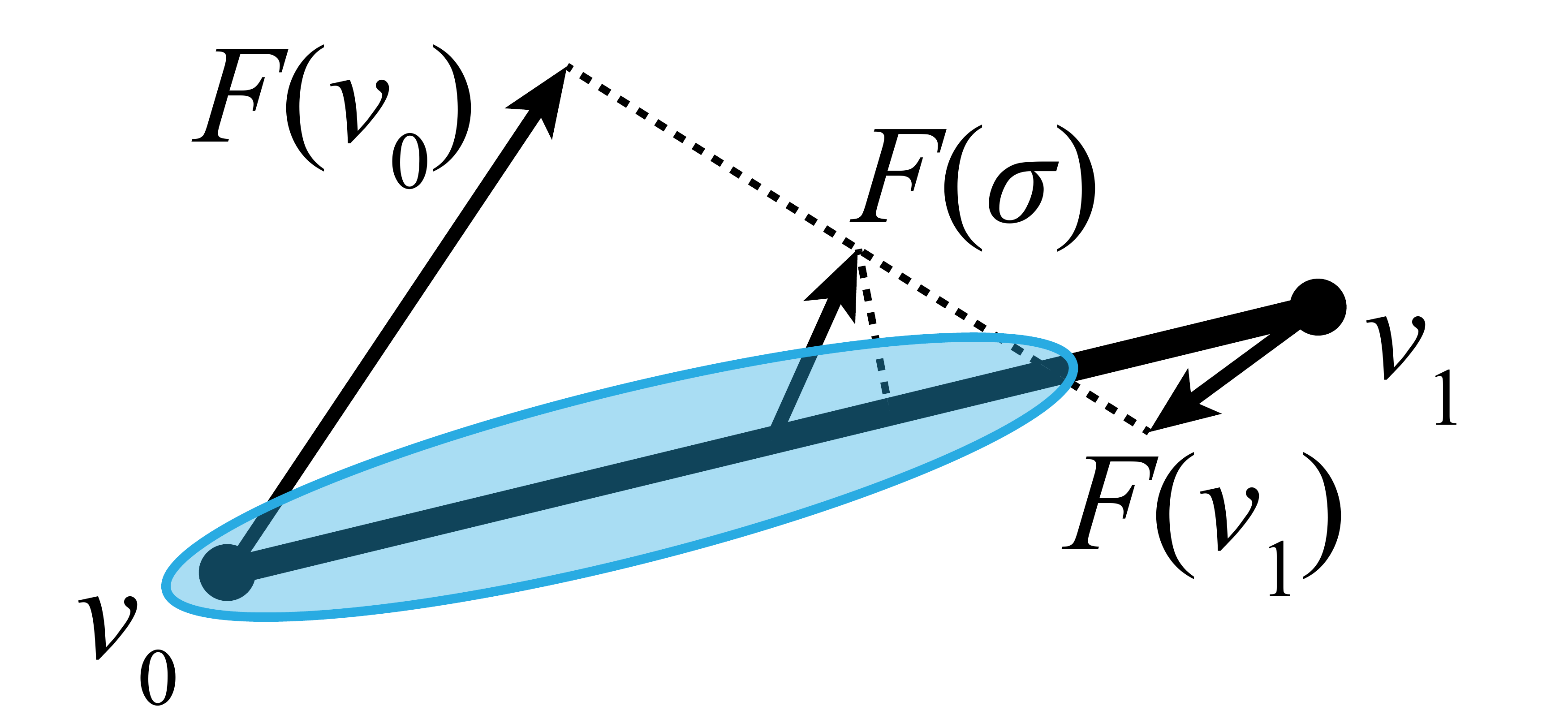}} 
      \vspace{-1em}
      \caption{}
      \vspace{-1em}
    \end{subfigure}
    \begin{subfigure}[b]{0.31\linewidth}
      \center{\includegraphics[width=\linewidth]{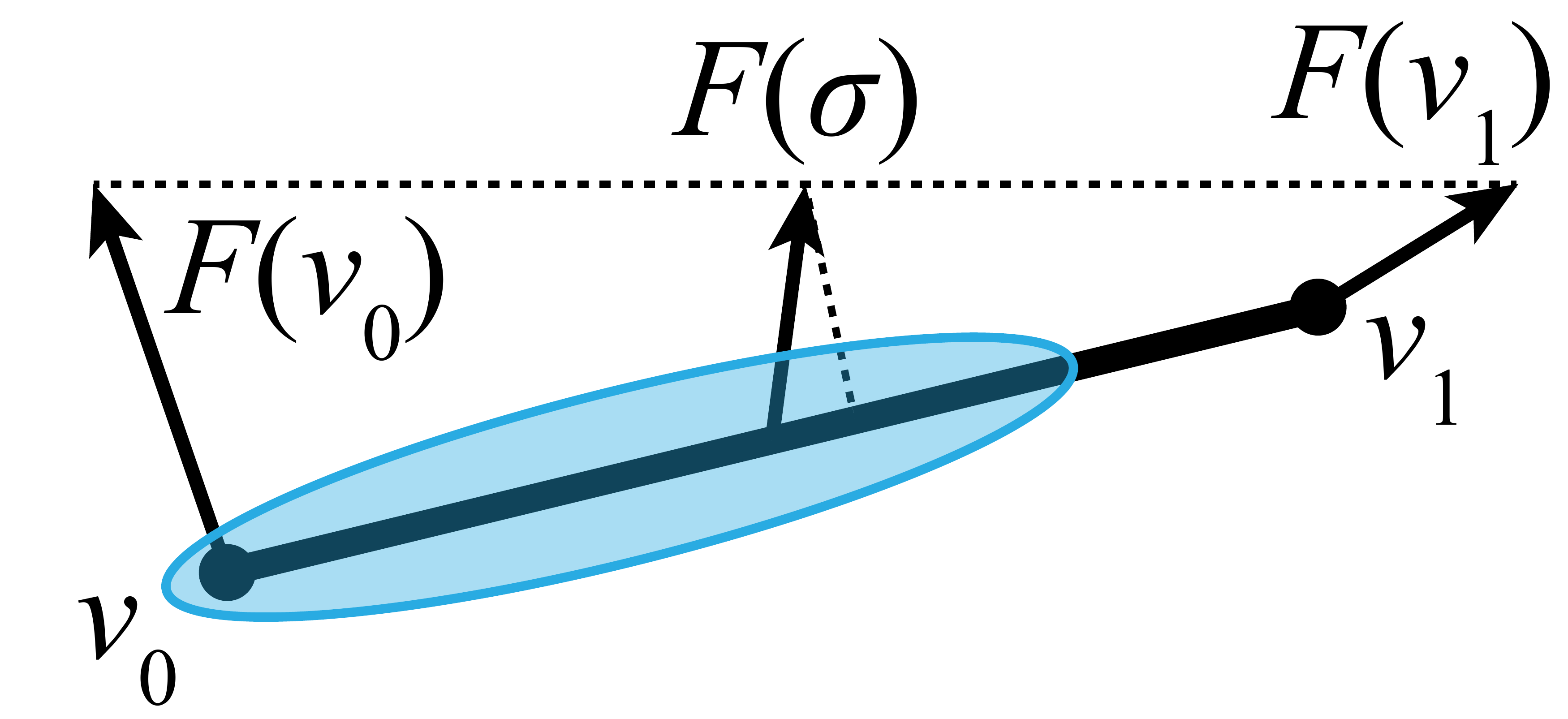}}
      \vspace{-1em}
      \caption{}
      \vspace{-1em}
    \end{subfigure}
    \caption{\label{fig:edgeStrength} We show three possible configurations for which $f(v_0,v_1)>0$ for a simplex $\sigma = \{v_0,v_1\}$, resulting in outward flow from $v_0$ (indicated by the blue ellipse.  Dotted lines indicate how $F(\sigma)$ and the dot product for $f(v_0,v_1)$ are computed.  In (a) both vectors are pointing away from $v_0$.  (b) The vectors at $v_0$ and $v_1$ both flow towards $\sigma$, but $F(v_0)$ dominates.  (c) Neither vector flows towards $\sigma$, however since $f(v_0,v_1)>0$ we define the flow as outward from $v_0$.}
\end{figure}

\begin{figure}[!t]
   \centering
   \begin{subfigure}[b]{0.18\linewidth}
      \center{\includegraphics[alt={A triangulated hexagon with various coloring to indicate flow from center vertex.},width=\linewidth]{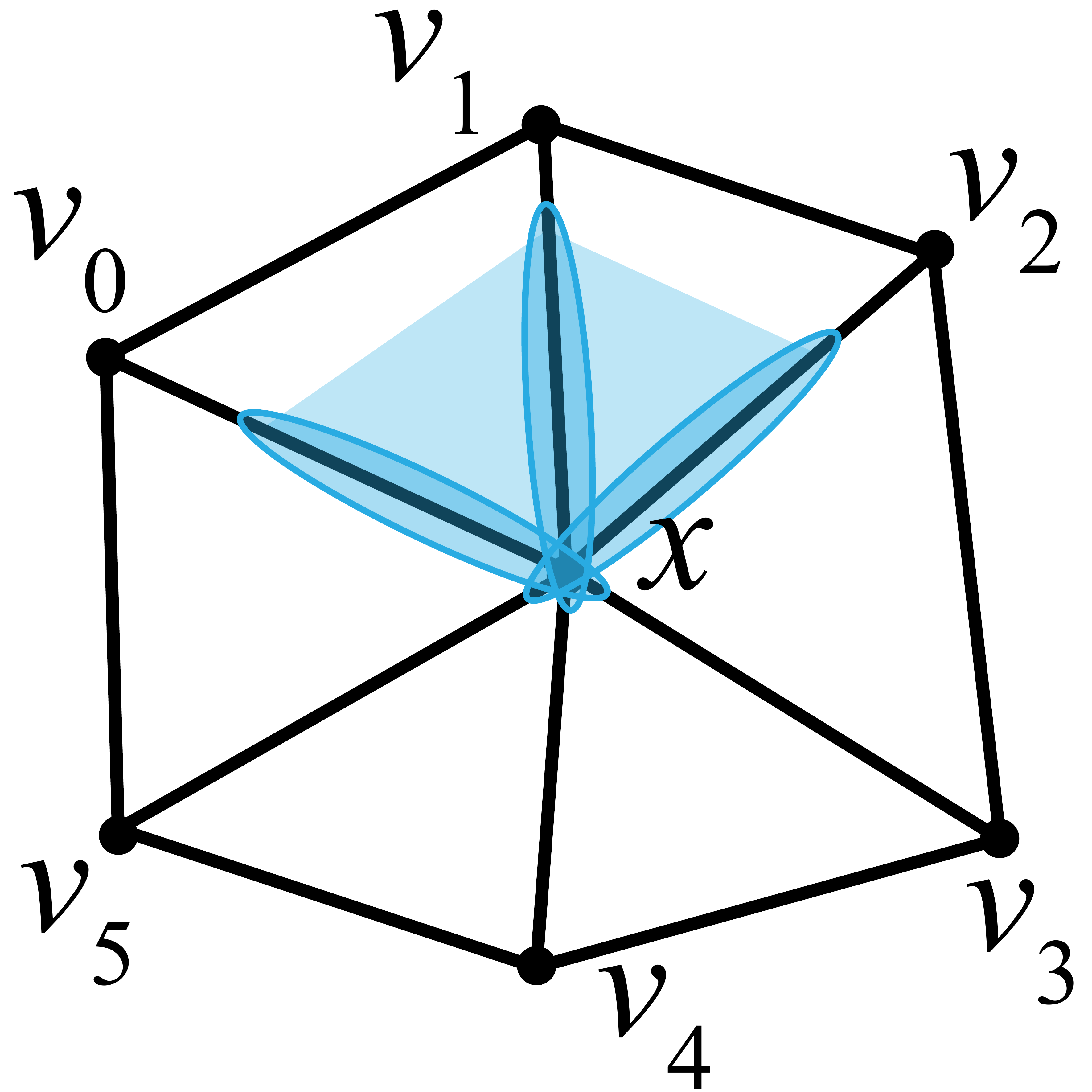}}
      \vspace{-1em}
      \caption{}
      \vspace{-1em}
   \end{subfigure}
   \begin{subfigure}[b]{0.18\linewidth}
      \center{\includegraphics[width=\linewidth]{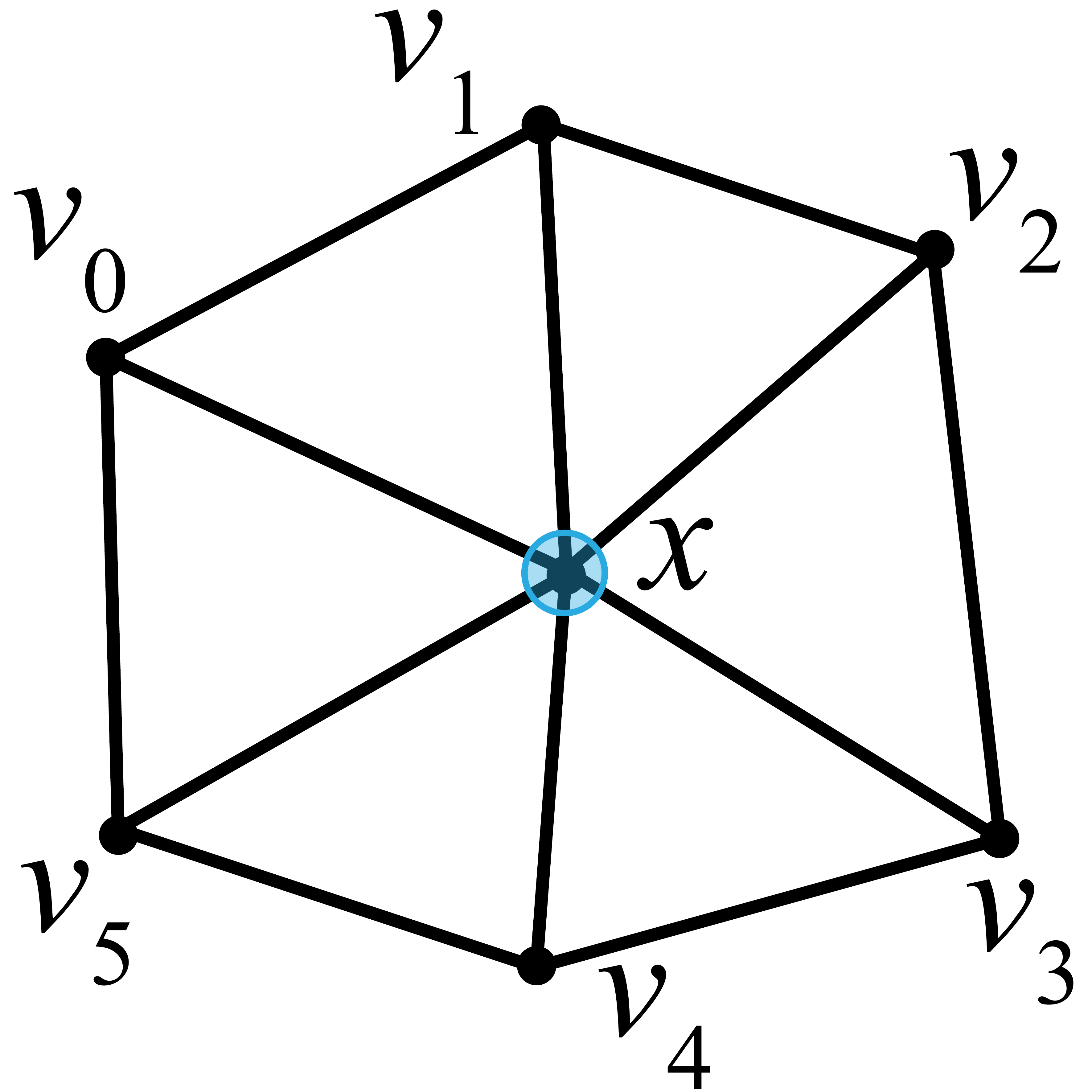}}
      \vspace{-1em}
      \caption{\label{fig:outstarZero}}
      \vspace{-1em}
   \end{subfigure}
   \begin{subfigure}[b]{0.18\linewidth}
      \center{\includegraphics[width=\linewidth]{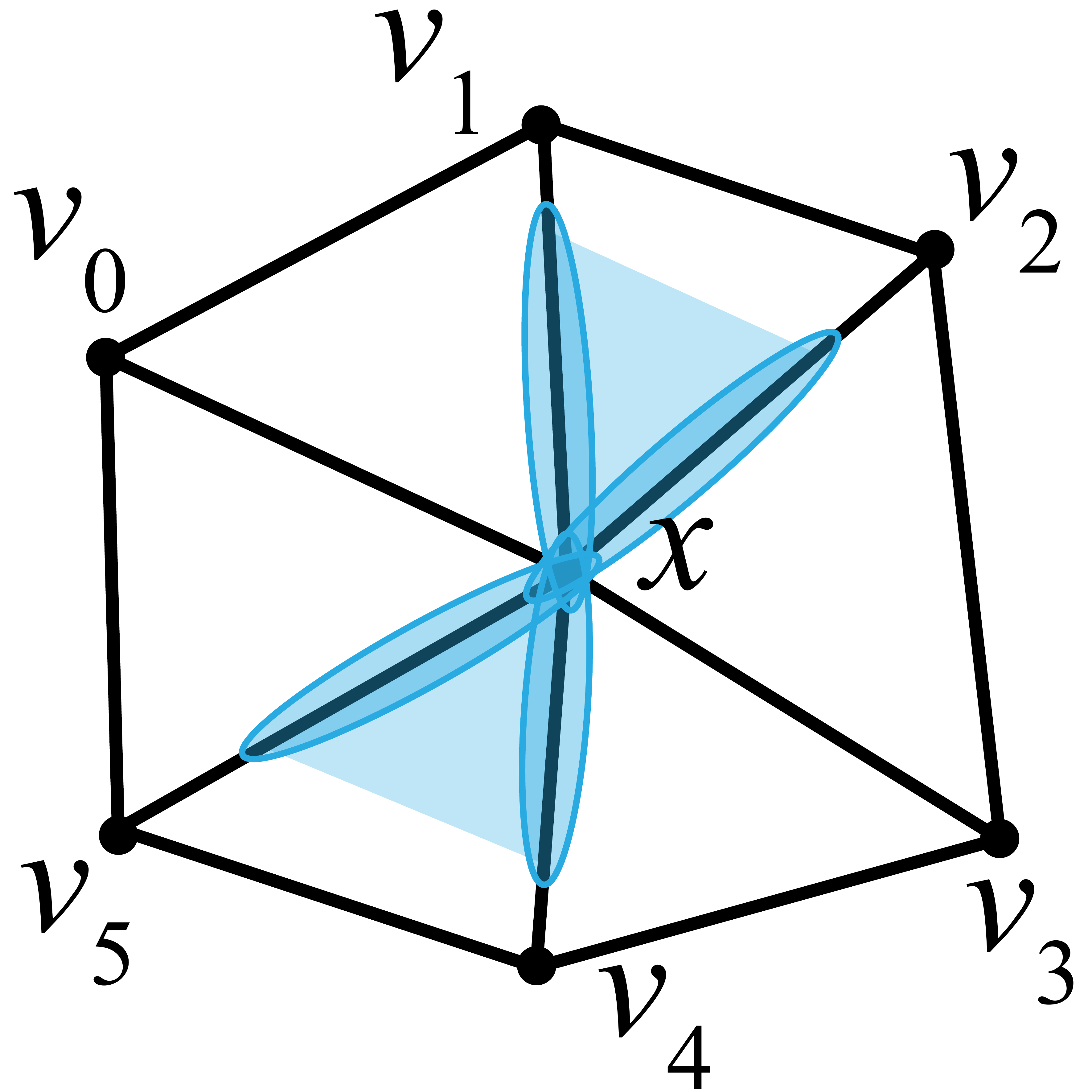}}
      \vspace{-1em}
      \caption{}
      \vspace{-1em}
   \end{subfigure}
   \begin{subfigure}[b]{0.18\linewidth}
      \center{\includegraphics[width=\linewidth]{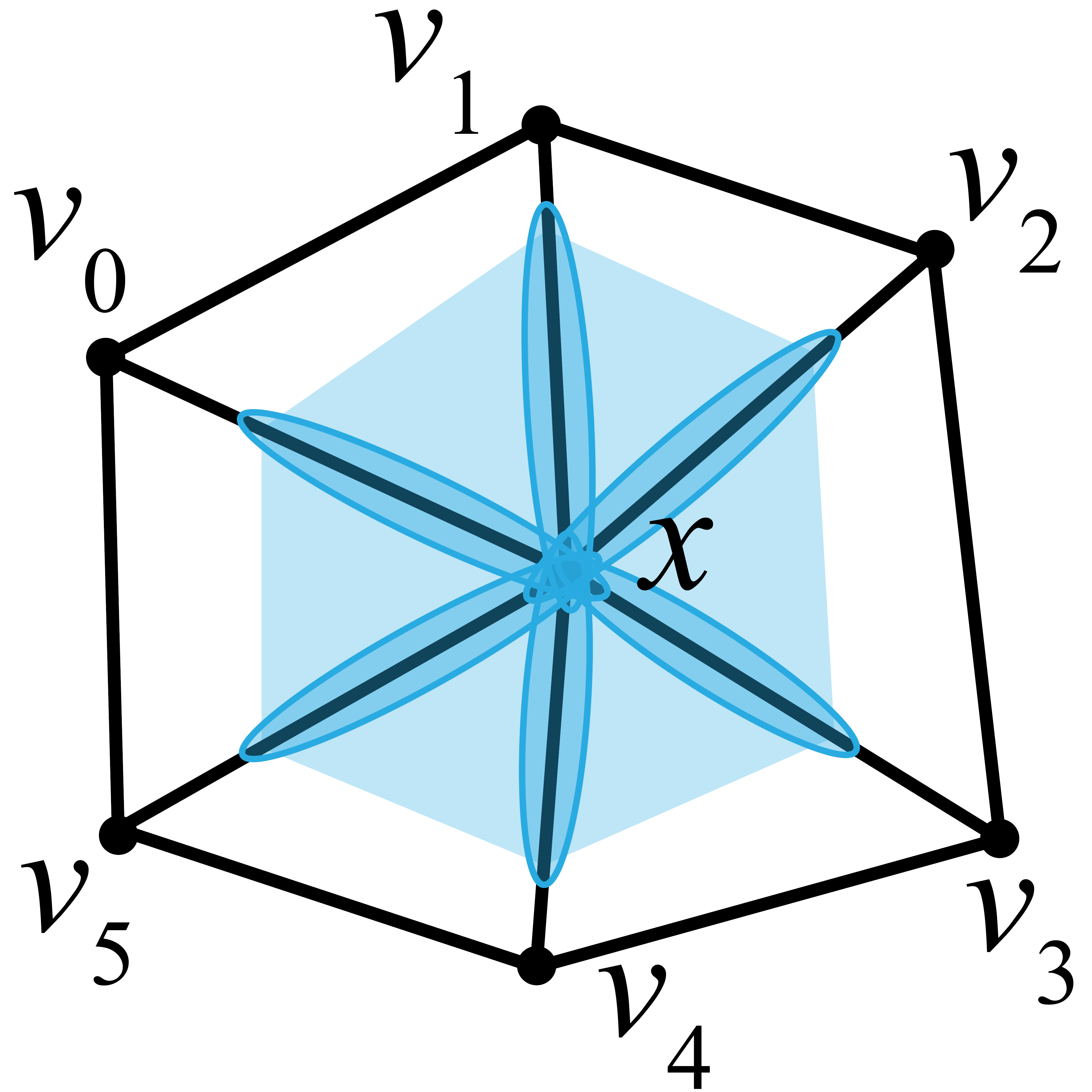}}
      \vspace{-1em}
      \caption{}
      \vspace{-1em}
   \end{subfigure}
   \begin{subfigure}[b]{0.18\linewidth}
      \center{\includegraphics[width=\linewidth]{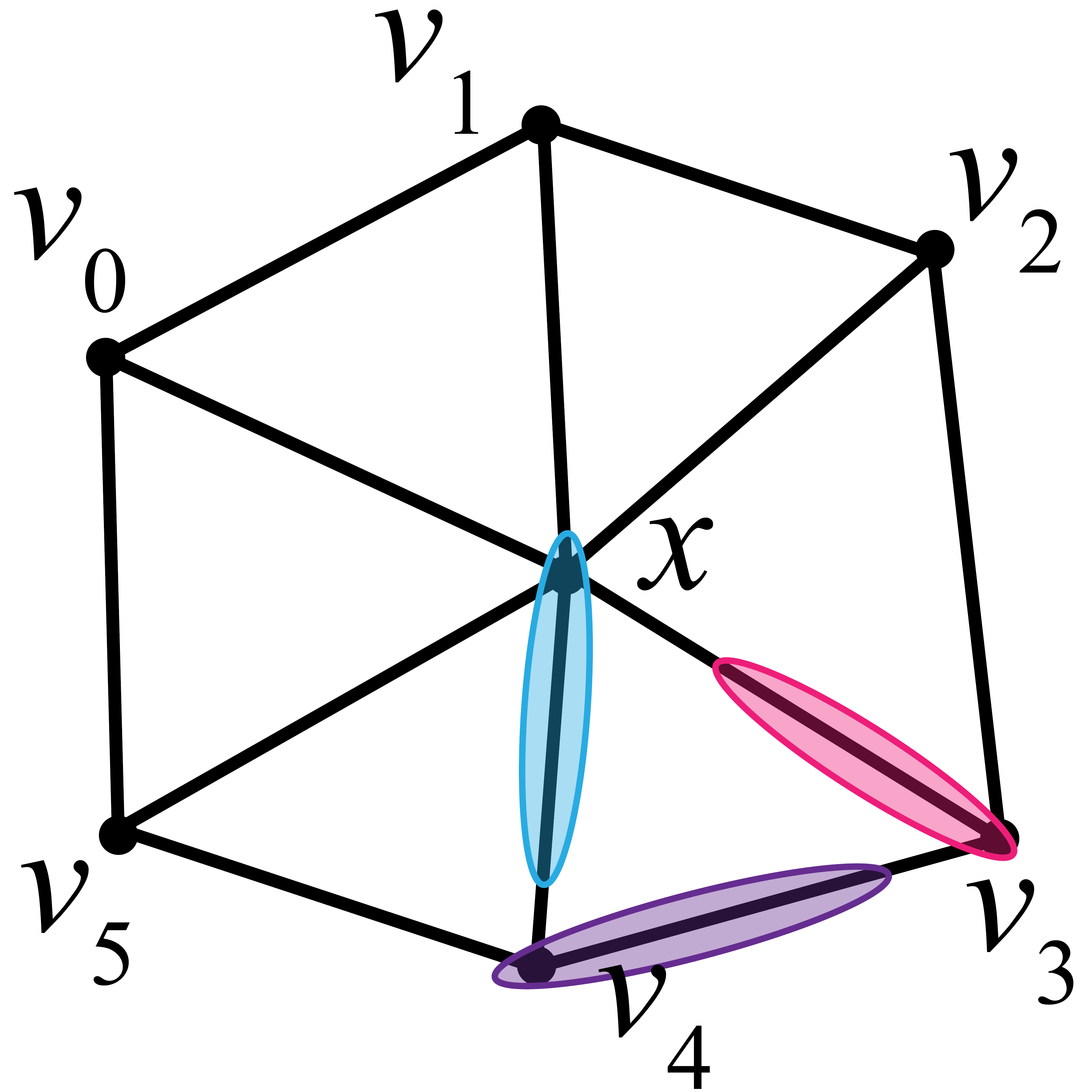}}
      \vspace{-1em}
      \caption{}
      \vspace{-1em}
   \end{subfigure}
   \caption{\label{fig:typeExamples} Examples of outward stars for a vertex $x$, \jaledit{with blue indicating} outward flow.  (a) $x$ has 3 edges (associated with $v_0,v_1,v_2$) and thus also includes two triangles in $\outStar(x)$. (b) No edges are in $\outStar(x)$.  (c) Four \jaledit{non-adjacent} edges are included in $\outStar(x)$, \jaledit{resulting in} two triangles \jaledit{in $\outStar(x)$} (involving $v_1, v_2$ and $v_4, v_5$). \change{(d) All surrounding edges and triangles are included in $\outStar(x)$}  (e) A lone edge is in $\outStar(x)$, involving $v_4$.  We also show edges assigned to $\outStar(v_3)$ (pink) and $\outStar(v_4)$ (purple).  In this case, the triangle will not be assigned to any outward star.}
\end{figure}

Our main goal is to develop a simple, consistent notion of outward flow for each vertex in a given  underlying triangulated mesh and vectors along each vertex provided $F(x)$. 
We focus first on 1-simplices since any choice made on a given edge can be used to propagate this definition to higher dimensional simplices.
Given an edge $\sigma$ with vertices $v_0, v_1$, one option is to consider weights that would result from pairing each $v_i$ to $\sigma$, \change{i.e.,}~$w(v_0, \sigma)$ and $w(v_1, \sigma)$.
However, it may be possible that both vertices experience a positive weight for pairing with $\sigma$, so to resolve this consistently (in the sense that both vertices agree on which direction has the dominant flow) we examine the weight of the $V$-path from $v_0$ to $v_1$.
If this $V$-path has positive weight, we describe the flow along $\sigma$ as flowing away from $v_0$.

While we could evaluate this directly with Eq.~\ref{eq:vpath}, it turns out that this equation can be more concisely written as 
\begin{eqnarray}
    \label{eq:edgeFlow}
    f(v_0, v_1) & = & w(v_0, \sigma) - w(\sigma, v_1) \nonumber \\
                & = & \frac{F(v_0) + F(v_1)}{2} \cdot (v_1 - v_0) \nonumber \\ 
                & = & F(\jaledit{c(\sigma)}) \cdot (v_1 - v_0) 
\end{eqnarray}
where $\sigma$ is the edge between $v_0$ and $v_1$.  
Interestingly, $f(v_0, v_1)$ captures both the direction (sign) and strength of flow from $v_0$ to $v_1$. 
If $f(v_0, v_1)$ is a positive value, we say that the edge $\sigma$ is \emph{outward flow}, while a negative value would be inward flow. 
Note that flipping the vertices results in negating the value, $f(v_0, v_1) = - f(v_1, v_0)$. 
Fig.~\ref{fig:edgeStrength} shows example configurations that result in assignments of outward flow, indicated by the blue ellipse.  

The outward star of a vertex $x$ is then defined using Eq.~\ref{eq:edgeFlow} on each of the simplices in $\St(x)$ to test if all the edges for that simplex have a positive notion of outward flow. 
We define the \emph{outward star}, $\outStar$, as follows:
\begin{equation}
    \label{eq:outStar}
     \outStar(x) = \{ \sigma \in \St(x) \ | \ \underset{v_i \in \sigma \setminus x}{\textrm{\large $\forall$}} f(x, v_i) > 0 \}
\end{equation}

We formed this notion to be consistent with a star so that if $\sigma \in \outStar(x)$ then any faces  $\tau < \sigma$ such that $\tau \in \St(x)$ will also have the property that $\tau \in \outStar(x)$.
This property is satisfied since we check all edges that are cofaces of $x$ and require that they all have a positive $f$ to include.
Said another way, a simplex of dimension greater than one is included in $\outStar(x)$ if all of its edges adjacent to $x$ are included.
This property ensures that we have a sufficient set of simplices in $\outStar(x)$ to perform a homotopic expansion.

Additionally, it is easy to see that the set of outward stars are all disjoint, as no edge will be in the outward star of more than one vertex.
Less obviously, unlike lower stars, outward stars will not form a partition of all simplices of the input mesh.
Under mild genericity assumptions\footnote{In the case of $f(v_0,v_1)=0$, neither vertex will request the edge.  We prefer to break ties in this case and assign the edge to the vertex of lower mesh id which ensures consistency.  We discuss alternatives in Section~\ref{sec:discussion}.}, all edges will have an obvious assignment to an outward star, but higher dimensional simplices might not be in any outward star. 
Fig.~\ref{fig:typeExamples} shows a number of examples of outward stars, and also includes an example where a triangle is excluded from any outward star due to all three of its edges being assigned to the outward stars of 3 different vertices.
In this case, a triangle will be excluded by processing from our algorithm and ultimately marked as critical, nevertheless this choice appears intuitive as such a triangle corresponds to a discrete equivalent of a center.


\subsection{Algorithm}

We next bring together these definitions to develop our algorithm for processing outward stars.

\begin{algorithm}[!t]
\caption{ProcessOutwardStars}\label{alg:1}
\hspace*{\algorithmicindent} \textbf{Input:} $K$ (Triangulated Mesh) \\
\hspace*{\algorithmicindent} \textbf{Input:} $F$ (Vector Field) \\
\hspace*{\algorithmicindent} \textbf{Output:} $V$ (Discrete Vector Representation)  $V[\alpha^p] = \beta^{p+1}$ \\
\hspace*{\algorithmicindent} \textbf{Output:} $C$ (Critical \jaledit{simplices}) 
\begin{algorithmic}[1]
\For{ $x^{(0)} \in K $ }
    \If{ \highlight{$\outStar(x)= \{ x \}$} }
        \State add $x$ to $C$ \Comment{x is a local sink} \label{alg:localSink}
    \Else 
        \State \highlight{ $\delta$ := the 1-\jaledit{simplex} 
        in $\outStar(x)$ s.t. $f(x,v_\delta)$ is maximal } \label{alg:edgeAssignment}
        \State $V[x]$ := $\delta $
        \State Add all other \highlight{$\beta^{(1)} \in \outStar(x)$} to PQzero
        \State Add all \highlight{$\alpha^{(2)} \in \outStar(x)$} to PQone s.t. $\alpha > \delta$ \newline
        \hspace*{4em} and numUnpairedFaces($\alpha$) = 1
        \While{PQone $\neq \emptyset$ or PQzero $\neq \emptyset$}
            \While{PQone $\neq \emptyset$} \label{alg:HEstart}
                \State $\alpha$ := PQone.popFront()
                \If{numUnpairedFaces($\alpha$)=0}
                    \State add $\alpha$ to PQzero
                \Else
                    \State $V[\textrm{pair}(\alpha)]$ := $\alpha$
                    \State remove pair($\alpha$) from PQzero
                    \State add all \jaledit{simplices} 
                    \highlight{$\beta \in \outStar(x)$} to PQone s.t. \newline
                    \hspace*{9em}($\beta > \alpha$ or $\beta > \textrm{pair}(\alpha)$) and \newline
                    \hspace*{9em}numUnpairedFaces($\beta$)=1
                \EndIf
            \EndWhile \label{alg:HEend}
            \If{PQzero $\neq \emptyset$} \label{alg:CPgenerate}
                \State $\gamma$ := PQzero.popFront()
                \State add $\gamma$ to C \label{alg:choseCP}
                \State add all \jaledit{simplices} 
                \highlight{$\alpha \in \outStar(x)$} to PQone s.t.\newline
                \hspace*{7em}($\alpha > \gamma$ and numUnpairedFaces($\alpha$)=1)
            \EndIf
        \EndWhile
    \EndIf
\EndFor
\State \highlight{Add all $\sigma \in K$ s.t. $\sigma \not\in \outStar(x)$ for all $x^{(0)} \in K$ to $C$} \label{alg:final}
\end{algorithmic}
\end{algorithm}

Algorithm \ref{alg:1} is largely a variation of Robins et al.'s ProcessLowerStars\cite{LowerStar} modified to work with an input piecewise linear vector field.  
We have highlighted in blue any lines that \change{were} significantly different from the original.
Specifically, one can see that we replace the concept of a lower star with an outward star.  
Line~\ref{alg:edgeAssignment} also selects the steepest edge to pair each vertex with, but replaces the search for the minimal adjacent vertex with a search for the edge $\delta = (x,v_\delta)$ which has a maximum weight according to Eq.~\ref{eq:edgeFlow}.

Like the original, the algorithm works by looping over each vertex $x$ of the mesh $K$, examining $\outStar(x)$.
Then assuming $v$ is not a sink (line~\ref{alg:localSink}), the process starts by determining a $\delta^1$ which has the most outward flow (equivalent to the steepest descent).
The edge chosen is paired with the vertex and then homotopy expansion occurs from lines \ref{alg:HEstart} to \ref{alg:HEend} with the help of priority queues PQone containing simplices with one unpaired face and PQzero containing simplices with zero unpaired faces. \change{A solely lexicographic order (where simplices are sorted just by steepest) can form loops in the remaining set, which may cause more simplices to go unpaired.}
We use pair($\alpha$) to refer to the single available unpaired face for the \jaledit{simplex} 
$\alpha$.
While any ordering in which a simplex is ranked after its faces will suffice~\cite{LowerStar}, we order our priority queues to reflect a notion similar to scalar field case.
Specifically, for a simplex $\sigma^{p} = (x,v_0,\ldots,v_{p-1})$, we look at all edges connected to $x$ and construct the negated edge weights $e_i = -f(x,v_i)$.
To capture steepest, we then construct the tuple in which the $e_i$'s are sorted in \change{decreasing} order, which is then sorted lexicographically. 
This both ensures that a simplex will appear after its faces, but processes the most outward simplices first. 
In practice, for two-dimensions (both scalar and vector fields), PQone rarely has more than one \change{or two items} in it at a time, and thus the choice on how to order the simplices plays out mostly in order of homotopy expansion.

After homotopy expansion, the if statement at line \ref{alg:CPgenerate} determines if a critical point is generated by the subsequent lines, choosing $\gamma$ to be critical (line \ref{alg:choseCP}) from the top of PQzero, then continuing homotopy expansion as necessary.  
Considering the configurations in Fig.~\ref{fig:typeExamples}, we can see that the set of items placed in the outer star will limit which candidates will be paired.
Configurations corresponding to Fig.\ref{fig:typeExamples}(b-d) will necessarily produce at least one critical simplex (of indices 0, 1, and 2, respectively) due to including an odd number of simplices to pair and the requirement that $x$ is always assigned to the steepest edge (line~\ref{alg:edgeAssignment}).
Fig.~\ref{fig:typeExamples}(e) \change{shows} an example of the type of configuring in which a 2-simplex may not be assigned to an outward star of any of its vertices.
Simplices not assigned to any outward star will never be checked as a candidate to pair, and thus we add them to $C$ (line~\ref{alg:final}).
In practice this detection happens implicitly by initializing all simplices to be unpaired (rather than requiring a final loop over all simplices).
As ties are broken for all edges, in two dimensions the only such simplices will also be index 2.

\subsection{Topological Simplification}

\begin{figure}[!t]
   \centering
   \begin{subfigure}[b]{0.32\linewidth}
      \center{\includegraphics[alt={A mesh with vectors along simplices of the mesh showing a saddle connected to 0- and 2-index critical points.},width=\linewidth]{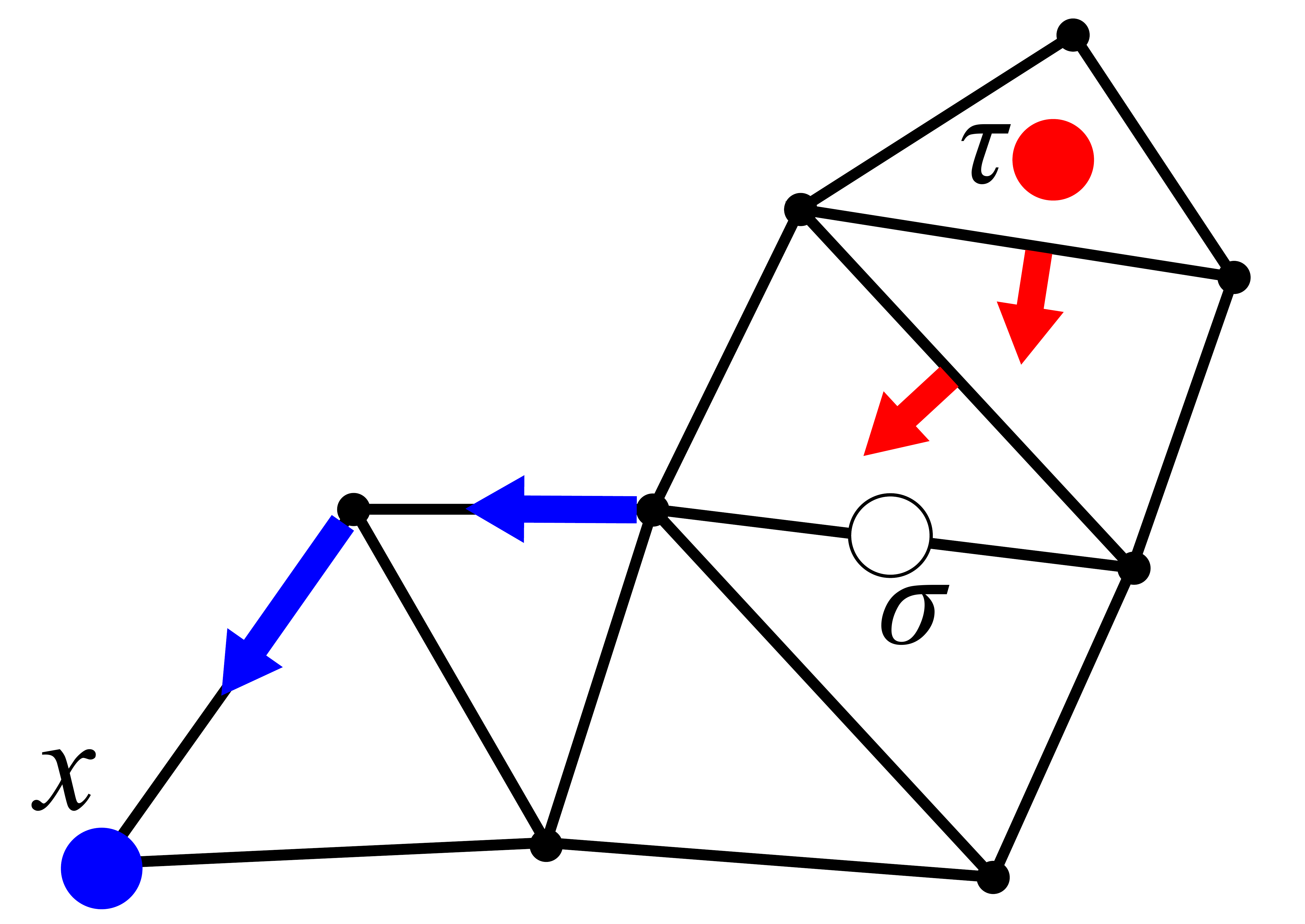}}
      \vspace{-1em}
      \caption{}
      \vspace{-1em}
   \end{subfigure}
   \begin{subfigure}[b]{0.32\linewidth}
      \center{\includegraphics[width=\linewidth]{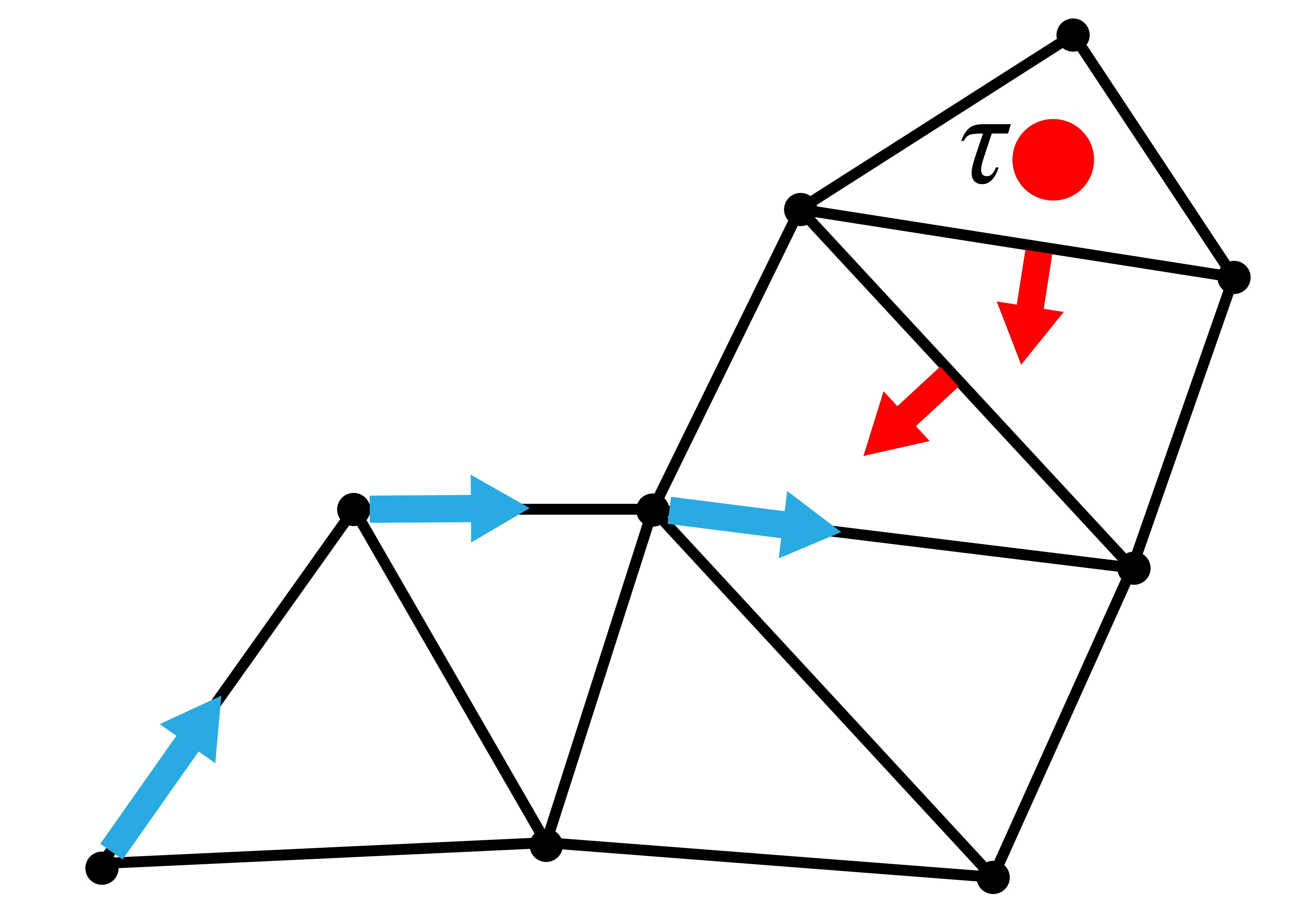}}
      \vspace{-1em}
      \caption{}
      \vspace{-1em}
   \end{subfigure}
   \begin{subfigure}[b]{0.32\linewidth}
      \center{\includegraphics[width=\linewidth]{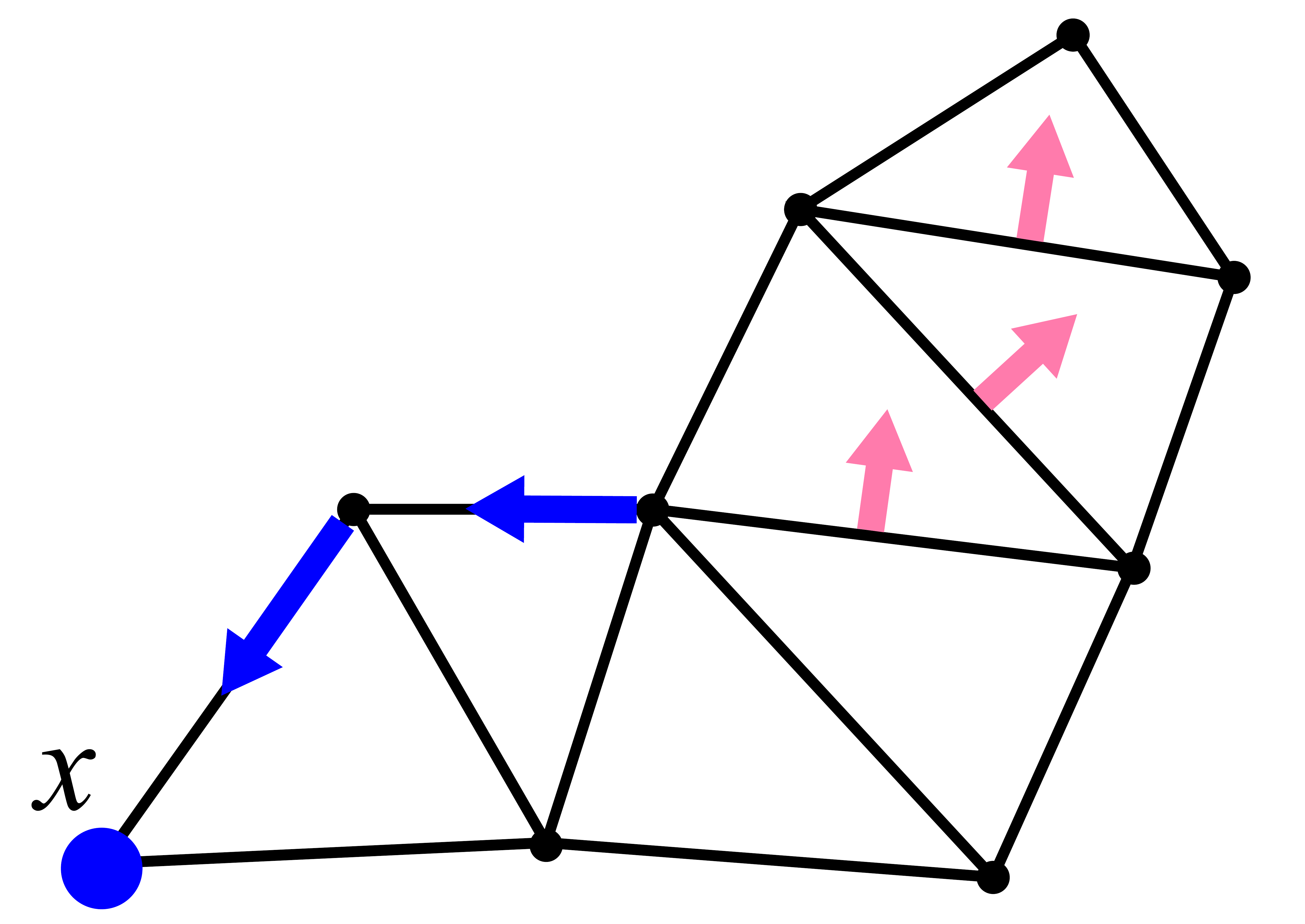}}
      \vspace{-1em}
      \caption{}
      \vspace{-1em}
   \end{subfigure}
   \caption{\label{fig:cancelSaddles} We show two different examples of saddle simplification. (a) in our input configuration, $\sigma$ has an index 0 separatrix that arrives at $x$ and an index 1 separatrix that arrives from $\tau$ (only two of the four possible separatrices are shown).  (b) Cancelling the index 0 separatrix results in removing both $x$ and $\sigma$.  (c) Cancelling the index 1 separatrix results in removing $\sigma$ and $\tau$.}
\end{figure}

Given a discrete vector field, not only is extracting topological structures straightforward, but the discrete representation enables a mechanism for simplifying the topology through cancelling critical points in pairs.
Any pair of critical points that are connected by a $V$-path can be cancelled through a process of $V$-path reversal~\cite{forman1998combinatorial}.
Specifically, reversing the $V$-path between critical simplices swaps the pairs and anti-pairs, which results in removing the critical simplices since they become part of a pair.
To identify candidate pairs, in two dimensions it suffices to use the four separatrices for each saddle.
We then can reverse the corresponding $V$-path simply removing  pairs ($\alpha_i,\beta_i$) and replacing them with the alternate pairs ($\beta_{i-1},\alpha_i$). Fig.~\ref{fig:cancelSaddles} shows an example of this for both index 0 and index 1 separatrices.

\begin{figure}[!t]
   \centering
   \begin{subfigure}[b]{0.32\linewidth}
      \vspace{-1em}
      \center{\includegraphics[alt={The same mesh as previous figure, but saddle connecting to two cycles formed by the vectors.},width=\linewidth]{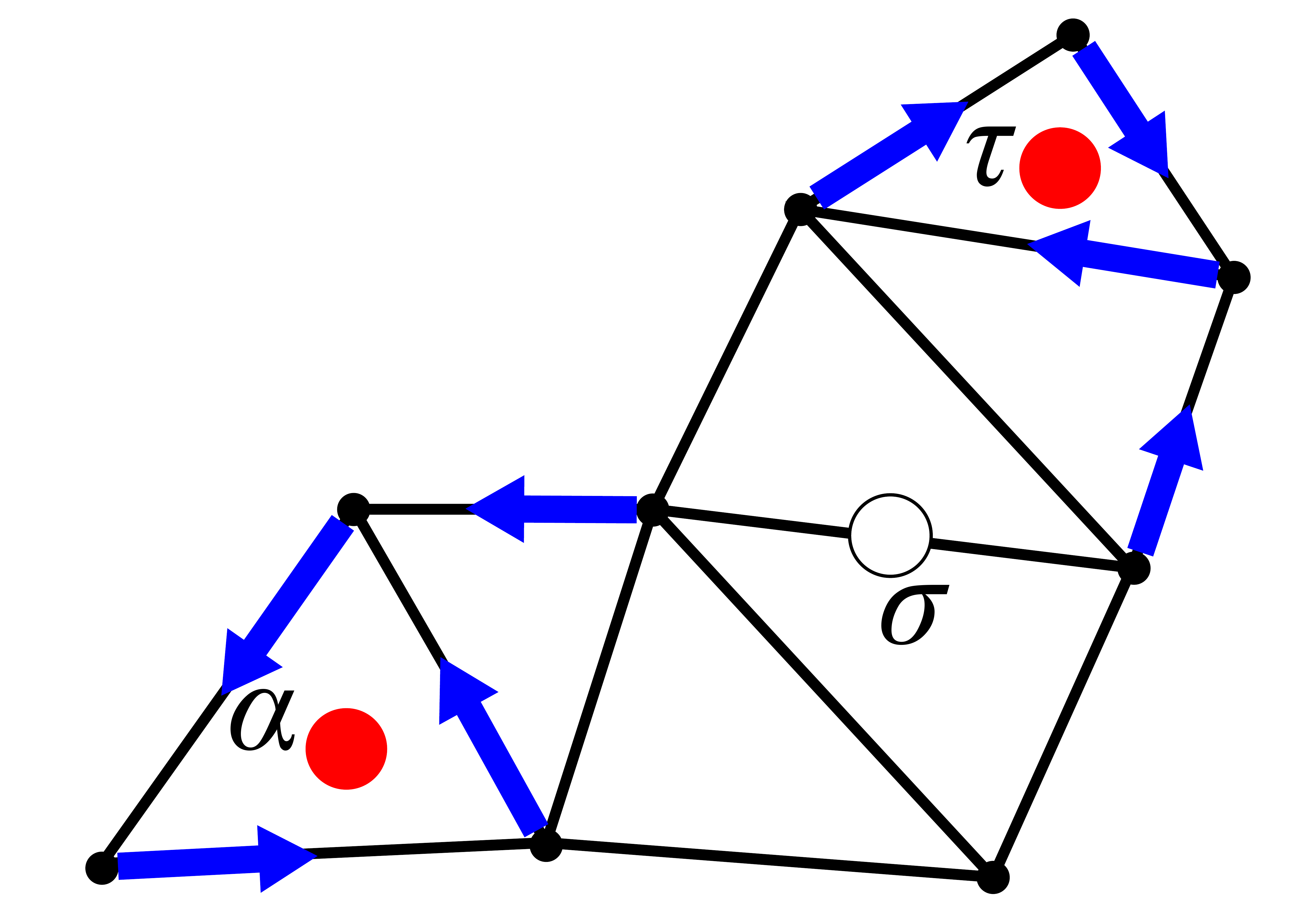}}
      \vspace{-1em}
      \caption{}
      \vspace{-1em}
   \end{subfigure}
   \begin{subfigure}[b]{0.32\linewidth}
      \vspace{-1em}
      \center{\includegraphics[width=\linewidth]{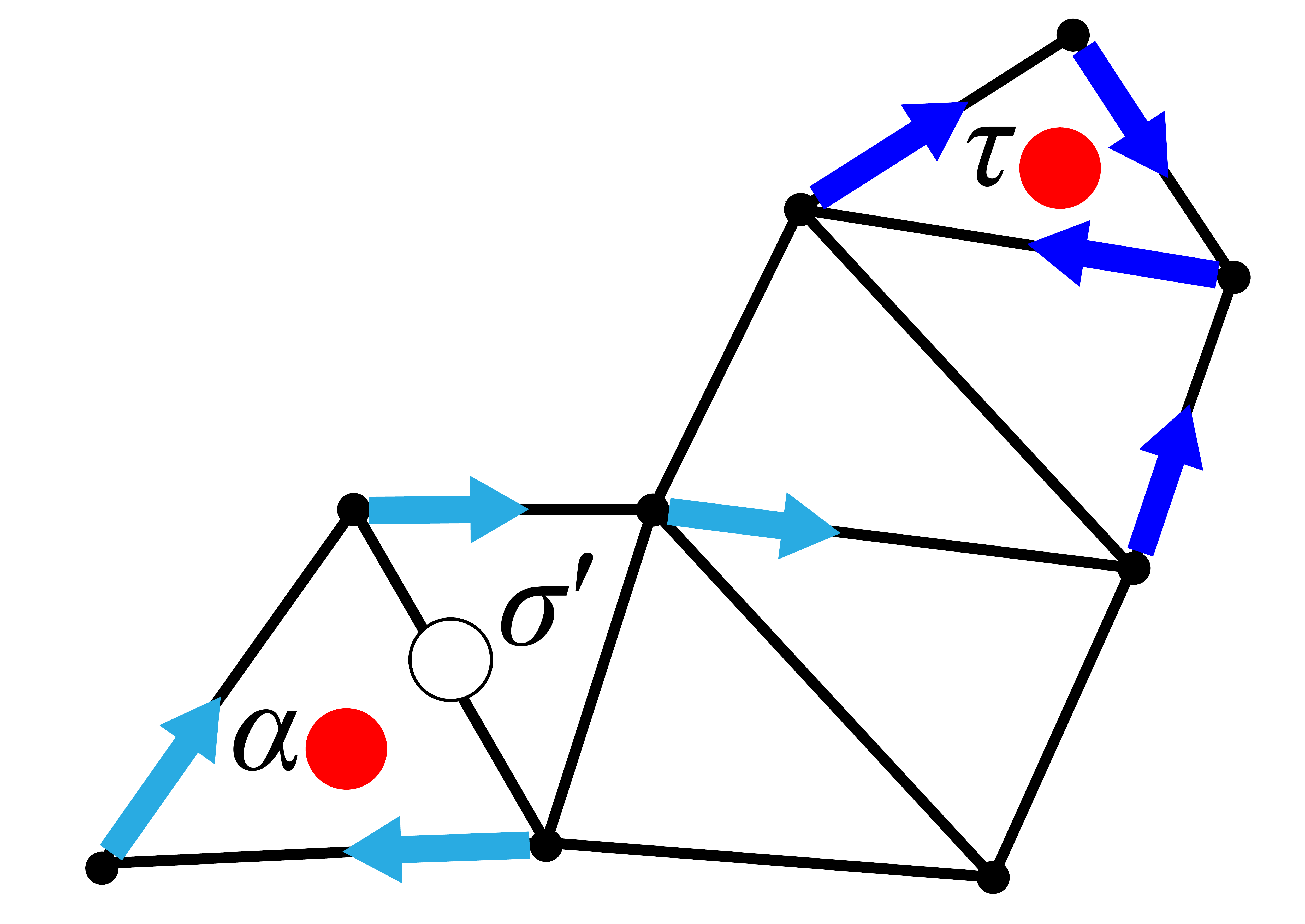}}
      \vspace{-1em}
      \caption{\label{fig:saddleLoop}}
      \vspace{-1em}
   \end{subfigure}
   \begin{subfigure}[b]{0.32\linewidth}
      \vspace{-1em}
      \center{\includegraphics[width=\linewidth]{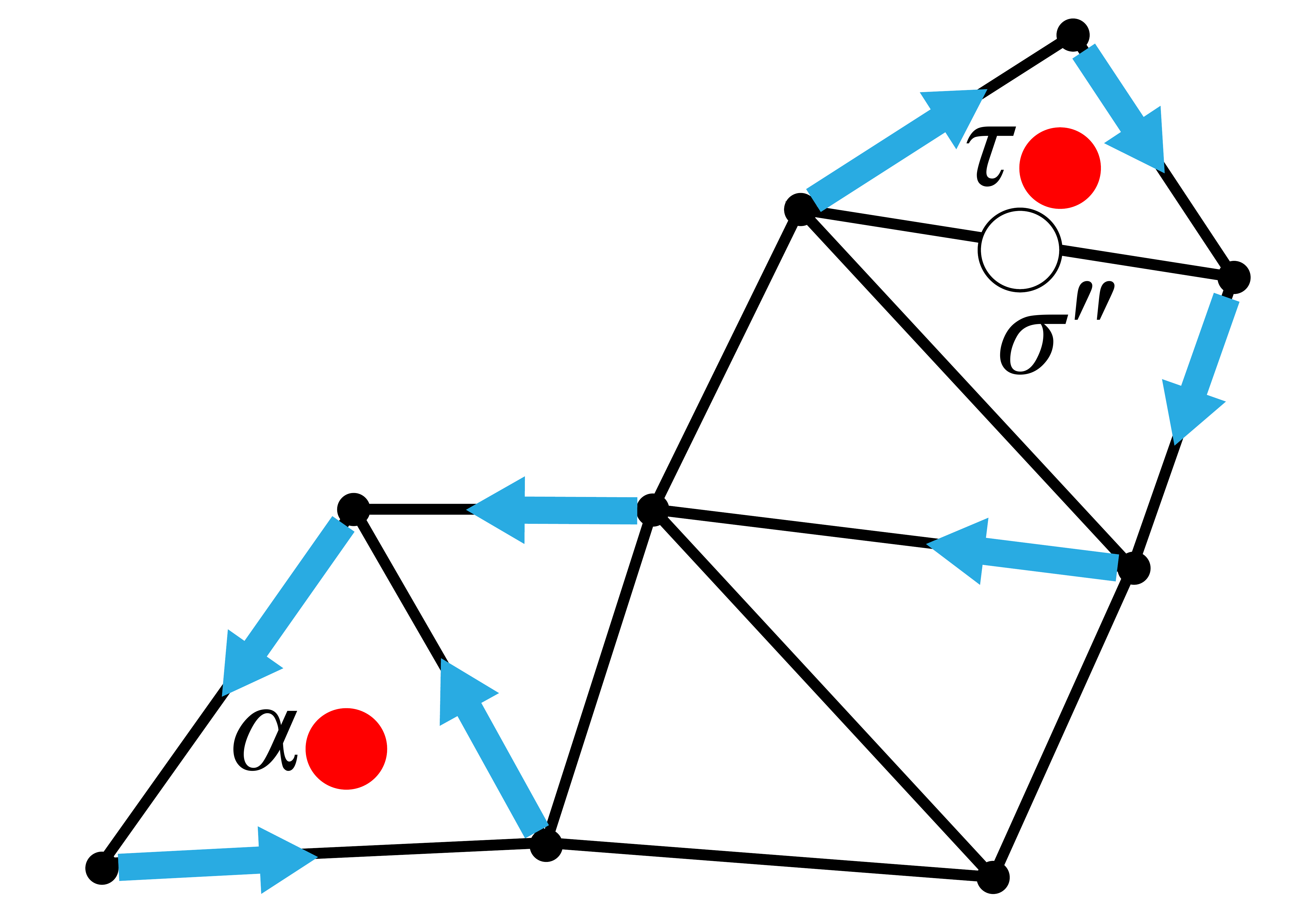}}
      \vspace{-1em}
      \caption{\label{fig:saddleLoopRepeat}}
      \vspace{-1em}
   \end{subfigure}
   \caption{\label{fig:edgeSimplification} Saddle-orbit cancellation.  In (a) we show a saddle $\sigma$ whose index 0 separatrixes both terminate in orbits.  Note for size we show an orbit on a single triangle, but a similar situation could occur with a longer closed $V$-path.  (b) Reversing the $V$-path that follows the orbit around $\alpha$ results in sliding $\sigma$ to $\sigma'$ and removing the orbit.  (c) If one tries to reverse the $V$-path that orbits $\tau$, $\sigma'$ will slide to $\sigma''$ and in doing so the orbit around $\tau$ will be removed, but the orbit around $\alpha$ will reappear.}
\end{figure}

Note that the equivalent algorithm for scalar fields could result in creating a closed orbit.  
We allow orbit creation in our approach as we are not limited to only produce gradient fields, although we did experiment with limiting cancellations to only those separatrices which would not create orbits.
That said, a couple of special cases need to be handled, described in Fig.~\ref{fig:edgeSimplification}.
First, some saddles will have separatrices that exit the boundary from the domain, these are excluded from this process as reversing them would not simplify the topology.
Second, some saddles have separatrices that approach a closed orbit rather than a critical point.  
In this case, reversing the $V$-path will not result in cancelling a pair of critical points, but this reversal will remove the orbit while sliding the saddle to be at the simplex along the $V$-path which is involved with the orbit (Fig.~\ref{fig:saddleLoop}). \change{Note that the final anti-pair in the cycle is not included because it can not be reversed as it would break the rule following discrete Morse theory that no simplex occurs in more than one pair. }
This \change{simplification} is conceptually similar to a homoclinic bifurcation described in the context of feature tracking in flow fields~\cite[Fig.~5]{tricoche2002topology}.
Finally, it may be the case that a saddle has two separatrices which approach orbits.  
In this case, cancelling one orbit will slide the saddle, but still maintain the second orbit. 
If one then tries to perform a cancellation of the second, reversal will result in the first orbit being recovered, while the second orbit being cancelled.
This process could repeat indefinitely (Fig.~\ref{fig:saddleLoopRepeat}), so we explicitly disallow saddle-orbit cancellations of index $i$ for saddles that have just cancelled an orbit of index $i$.

In addition, for this approach to work we need to decide which $V$-paths to reverse and in what order.
For scalar fields, the above cancellation process could utilize relative change in function value to define persistence~\cite{edelsbrunner02} as a measure to rank features.
Trichoche et al.~have defined similar measures for ranking separatrices of vector fields, focusing on changes in vector magnitude and length of paths~\cite{tricoche2000topology,tricoche2001continuous}.
As the weight of a $V$-path (Eq.~\ref{eq:vpath}) accumulates a similar concept, we employ this to rank separatrices, choosing the separatrix with the least weight to cancel first.
Note that in general, while not all separatrices of a gradient field will be persistence pairs, in two dimensional scalar fields the separatrix with the smallest height difference will always correspond to a persistence pair.
Thus, after each cancellation, we update existing $V$-paths that were adjacent to the cancelled pair of critical simplices, and then reevaluate the set of separatrices to identify the next minimum weight separatrix.

Algorithm~\ref{alg:simplify} summarizes this approach, which takes as input the discrete vector field $V$, a set of critical simplices $C$, and a target number of saddles $N_s$.  
We use the notation $C^{(1)}$ to refer to the index 1 subset of $C$, \change{i.e.,} the critical 1 simplices corresponding to saddles.
We use the routine FollowVPaths to follow the V-Paths along each of the saddles (line \ref{alg:initVpaths}) and produce the priority queue, PQpairs, of sorted possible pair reductions.
Pairs are simplified by reversing the discrete vectors as described, and then we update the existing set of pairs in our priority queue.
To accelerate this, we also maintain a bookkeeping structure, Connections, which keeps track of all connecting extrema from a given saddle.
This process is repeated until the number of saddles in the field is below $N_s$.

\begin{algorithm}[!t]
\caption{Saddle Based Simplification Algorithm}\label{alg:simplify}
\hspace*{\algorithmicindent} \textbf{Input:}  $V, C, N_s$ \\
\hspace*{\algorithmicindent} \textbf{Output:} $V, C$ s.t. $|C^{(1)}| = N_s$ \label{alg:outputDescription}
\begin{algorithmic}[1]
\State PQpairs, Connections := FollowVPaths($V,C^{(1)}$) \label{alg:initVpaths}
\While{$|C^{(1)}| > N_s$ and PQpairs $\neq \emptyset$}
    \State $(\alpha,\beta)$ = PQpairs.popFront()
    \If{ $\alpha \in C$ and ($\beta \in C$ or $\beta$=orbit)} \Comment{$\alpha, \beta$ not cancelled.}
        \State ReversePairs($V$,$C^{(1)}$,$(\alpha,\beta)$) \label{alg:flipLinks}
        \State UpdatePairs($V$,$C^{(1)}$,$(\alpha,\beta)$,PQpairs, Connections) \label{alg:updatePaths}
    \EndIf
\EndWhile
\end{algorithmic}
\end{algorithm}


By comparison, Reininghaus et al.'s FastCVT approach~\cite{fastCVT} to constructing a discrete vector field has a natural simplification mechanism built in (with minor modifications, one can set a threshold for number of critical points and optimize the matching globally to produce a field with a desired number of critical points).
We utilize this method for comparison in Section~\ref{sec:exp}.
In our case, we also considered starting with the discrete vector field produced by running Algorithm~\ref{alg:1} and then applying the simplification optimization of Reininghaus et al.
Note however that this is similarly expensive to computing a target vector field with FastCVT (\change{i.e.,} requires multiple iterations of Bellman-Ford), and moreover it is also more expensive than our saddle based approach.
In addition, it also assumes\cite[Sec.~6]{fastCVT} that there are no negative weighted cycles in the shortest path search of the bipartite graph equivalent to the discrete vector field.
A cycle with negative weight leads to a potential infinite loop, as traversing a negatively weighted cycle will result in situation where this is no shortest path (as repeated walks along the cycle will continue to look cheaper).
Our algorithm does not explicitly prevent such cycles from forming, but we also do not need a shortest path computation to make decisions on how to build discrete vectors nor to simplify.

\section{Experimental Evaluation\label{sec:exp}}
This section presents a review of the results obtained from running our algorithm, then comparing it to FastCVT~\cite{fastCVT} and the continuous case.
Unless otherwise noted, the results for FastCVT only include extraction of critical points based on the notion of maximum weighted global choices. Where applicable we include timing for simplification until a 0 threshold. In FastCVT, this approach is equivalent to the complete simplification and then rolling back the simplification until the max weighted discrete vector field. 

Experiments are run on a machine running Ubuntu 22.04, an Intel(R) Core i7-9700K CPU @ 3.60GHz with 64GB RAM. Our algorithm and FastCVT were both implemented in C++ as a TTK filter for triangulated vector fields~\cite{TiernyFLGM18}, and our experiments were powered by ParaView~\cite{paraviewBook}.  Our own version of FastCVT (based on the original algorithm~\cite{fastCVT}) was also implemented in TTK to provide an equal comparison pipeline, although timings compared with the original may vary.

\begin{table*}[htbp]
    \centering
    \caption{Runtime Comparison of Extraction using Outward Star Algorithm (OSA) versus our implementation of FastCVT\cite{fastCVT} on different datasets with differing number of critical points produced. For each dataset we report the total number of simplices ($|K|$) and the total number of critical simplices ($|C|$) produced by each algorithm. Also the associated time to extract and simplify (\textbf{Simp.}) to the \textbf{Target} number of critical points representing full simplification of that field in the last two columns. Note: full simplification is sometimes >1 because of border artifacts in the mesh. Exceptions: The symbol $\dag$ indicates it took an unreasonably long time to run and was stopped early.} 
    \label{tab:timingComparison} 
    \begin{tabular}{|l||r|r|r|r|r|r|r|r|}
        \hline
        \textbf{Dataset} & \textbf{$|K|$} & \textbf{OSA (s)} & \textbf{$|C|$, OSA} & \textbf{FastCVT (s)} & \textbf{$|C|$, FastCVT} & \textbf{Target $|C|$} & \textbf{OSA Simp. (s)} & \textbf{FastCVT Simp. (s)} \\
        \hline
        \hline
        Cosine & 350K & 1.580 & 37 & 25.014 & 39 & 1 & 2.947 & 57.687  \\
        \hline 
        \change{Changes} & 538K & 2.154 & 27 &  35.664 &  33 & 1 & 2.320 & 168.319 \\ 
        \hline
        RT70 & 716K & 3.301 & 367 & 147.074 & 315 & 1 & 6.120 & 899.056 \\ 
        \hline
        Ocean & 1M & 4.528 & 1,577 &  307.583 & 1,383 & 103 & 13.705 & 952.395 \\
        \hline
        OceanBig & 48M & 239.161 & 65,376 &  >$55000^\dag$ & $\dag$ & 1,340 & 691.975 & >$55000^\dag$\\
        \hline
    \end{tabular}
    \vspace{-2ex}
\end{table*}

We experiment with a collection of datasets in Table~\ref{tab:timingComparison}.  Specifically, the 
\jaledit{Changes and Cosine datasets} are the result of analytic equations described below.  
\jaledit{RT70 was created with Basilisk~\cite{popinet2013basilisk}. }
RT70 is the 70th time step of a Rayleigh-Taylor instability\footnote{\url{http://basilisk.fr/sandbox/Antoonvh/rt.c}}, cropped 
\jaledit{at the center.}
Ocean is daily mean surface velocity of the ocean from January 8, 1992, 
\jaledit{from}
the ECCO Consortium\cite{ECCO}. OceanBig is from the CMIP6 Earth Systems Grid Federation, specifically examining the monthly average of the ocean velocity components for January, 2014 on the surface of the ocean for the HadGEM3-GC31-HH climate model~\cite{CMIP6}.

Table~\ref{tab:timingComparison} shows timings of various datasets run using our algorithm compared to running FastCVT along with complete simplification times. The largest dataset, OceanBig, was run for over 14 hours and did not finish processing with FastCVT (as indicated with **).  


\subsection{Comparisons}

In order to show our algorithm's similarity to the piecewise linear case, we compared to the piecewise linear topological skeleton, implemented through the VTK Vector Field Topology filter\cite{vtkFilter}. 
First, we show how our algorithm can accurately identify critical points from simple cases of vector field flow, then progress to showing more complex datasets that exhibit additional behaviors and necessitate simplification. 

\begin{figure}[!t]
   \centering
   \vspace{-1ex}
   \begin{subfigure}[t]{0.49\linewidth}
      \includegraphics[alt={Side by side comparison of spheres representing critical points on the underlying flow.},width=\linewidth]{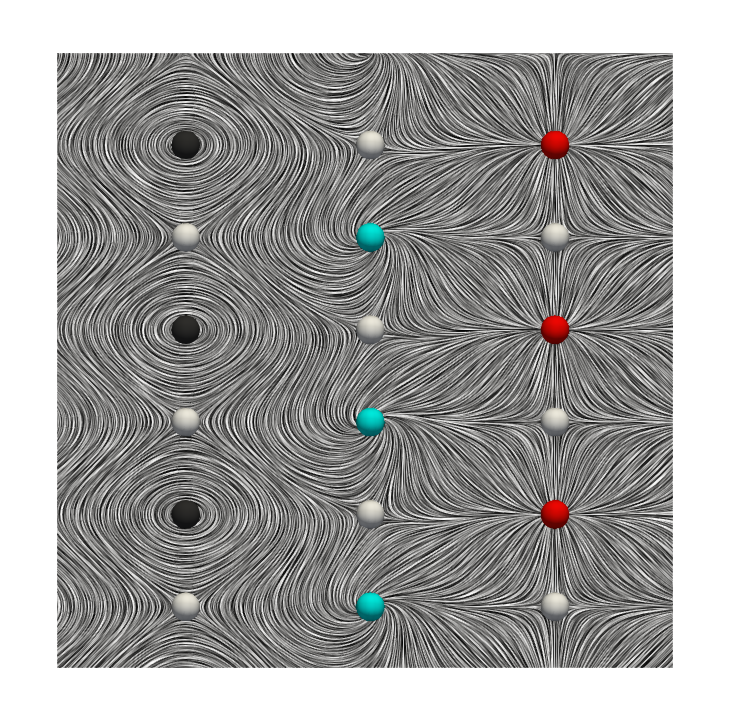}
      \vspace{-2em} 
      \caption{\label{fig:simpleContinuous}PL}
      \vspace{-1em} 
   \end{subfigure}
   \begin{subfigure}[t]{0.48\linewidth}
      \includegraphics[width=\linewidth]{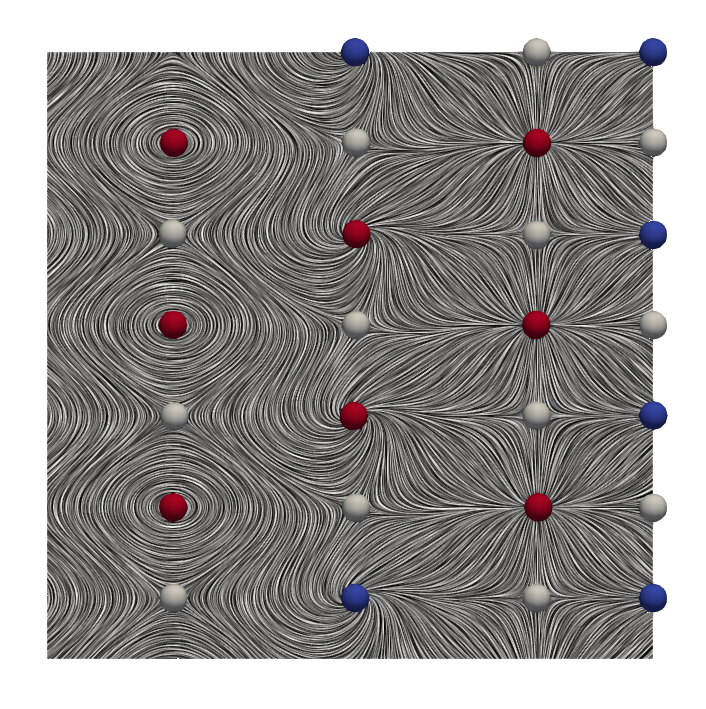}
      \vspace{-2em} 
      \caption{\label{fig:simpleResult}Discrete}
      \vspace{-1em} 
   \end{subfigure}
    \caption{ Comparison of Vector Field produced by Formula \ref{eq:vectorField} (a) Continuous evaluation using (Colors: 
    Cyan=Attracting Focus, White=Saddle, \change{Dark-Grey=Center,Orange=Repelling Node})  (b) Discrete Vector Field critical points produced by our algorithm 
    (Discrete \change{Sphere} Colors: Red=Index 2 CP, White=Index 1 CP, Blue=Index 0 CP. ) Additional discrete critical points \change{in upper right are} due to boundary artifacts.
    \label{fig:simpleFigure}
    }
\end{figure}

We generated a vector field using a formula that produces cycles\change{, attracting foci, and repelling nodes across the $x$-axis. }
This dataset, \change{called Changes}, is \jaledit{defined by} Eq.~\ref{eq:vectorField} \change{which combines two vector fields which are 90 degree rotations of each other}.  
We sampled a domain $x,y \in \change{[0,300]}$ uniformly every \change{1} units in both the $x$ and $y$ directions.
\change{The vector field produced has 18 continuous critical points as displayed in Fig.~\ref{fig:simpleContinuous}. Our algorithm produces a total of 27 discrete critical points \jaledit{(Fig.~\ref{fig:simpleResult})} capturing the 18 continuous critical points, and \jaledit{also having 9 critical points on the boundary.}}

\begin{equation}
\begin{split}
\label{eq:vectorField}
\change{f(x, y) =
    \begin{cases} 
        R(100)  
        \begin{pmatrix}
        \sin(a(y)) \\
        \cos(b(x))
        \end{pmatrix}
        = \begin{pmatrix}
        \sin(a(y)) \\
        \cos(b(x))
        \end{pmatrix} & x \in [0,100) \\
        R(x)  
        \begin{pmatrix}
        \sin(a(y)) \\
        \cos(b(x))
        \end{pmatrix} & 
          x \in [100,200) \\
        R(200)  
        \begin{pmatrix}
        \sin(a(y)) \\
        \cos(b(x))
        \end{pmatrix}
        = \begin{pmatrix}
        -\cos(b(x)) \\
        \sin(a(y))
        \end{pmatrix} &  x \in [200,300]
    \end{cases}} 
\end{split}
\end{equation}
\change{where $b(x) = 0.035(x - 17.5)$, $a(y) = 0.07(y + 15)$, and}
\[
\change{
\theta(x) = \frac{x - 100}{100} \cdot \frac{\pi}{2}, \quad 
R(x) = \begin{pmatrix}
         \cos(\theta(x)) & -\sin(\theta(x)) \\
         \sin(\theta(x)) & \cos(\theta(x))
        \end{pmatrix}.
}
\]

\jaledit{We designed this example with multiple considerations in mind.  As shown in Fig.~\ref{fig:simpleFigure}, this field exhibits a mix of vector field types, being purely incompressible on the first (leftmost) region, blending compressible and incompressible in second (center) region, and exhibiting irrotational flow on the third (rightmost) region.   As such, the flow on the left is divergence free and the flow on the right is rotational free.  We anticipated the third region to behave similar to scalar field topology, as it is like a gradient field.  For the center and left regions, we see that discrete critical points manifest in similar positions, and while they have similar indices, discrete flow can vary their type.}



\begin{figure}[!t]
   \centering
   \begin{subfigure}[t]{0.49\linewidth}
      \includegraphics[alt={Another example of flow with discrete critical points and separatrice lines shown.},width=\linewidth]{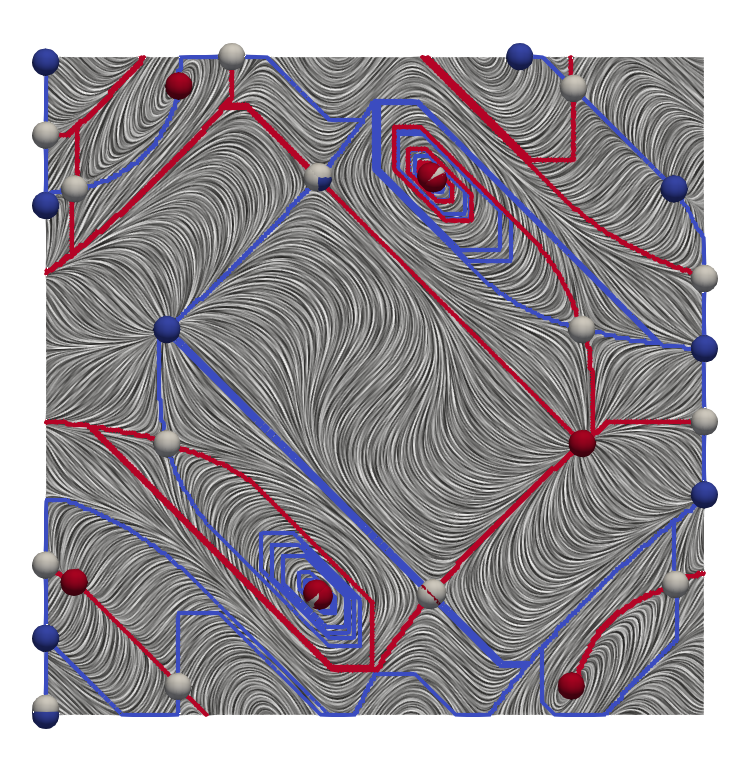} 
      \vspace{-2em}
      \caption{\label{fig:cosineOurs}OSA}
      \vspace{-1em} 
   \end{subfigure}
   \hfill
   \begin{subfigure}[t]{0.49\linewidth}
      \includegraphics[width=\linewidth]{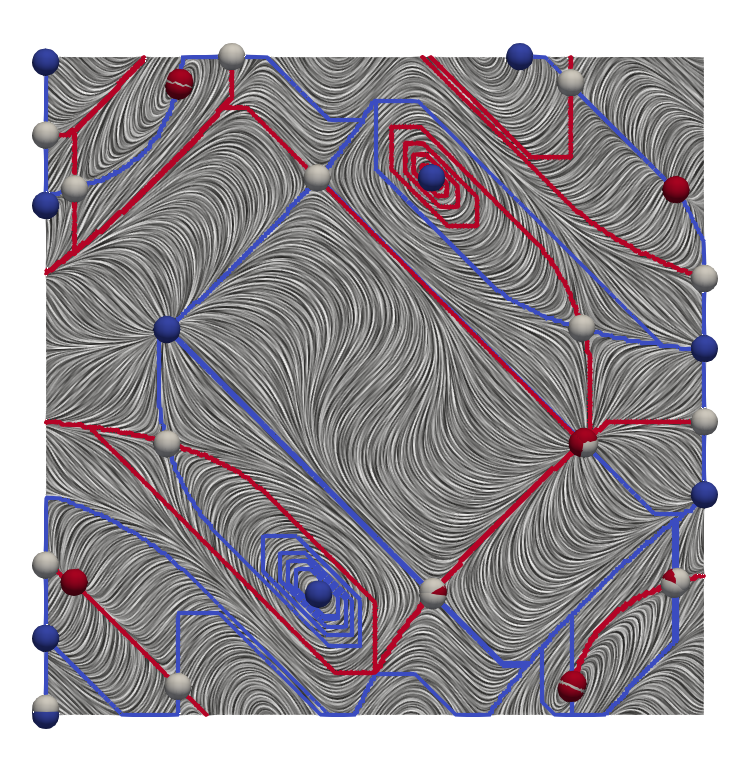}
      \vspace{-2em}
      \caption{\label{fig:cosineFast}FastCVT}
      \vspace{-1em} 
   \end{subfigure}
    \caption{Cosine: Discrete Vector Field Extraction Comparison (a) produced by our algorithm (OSA) (b) produced by FastCVT. Streamlines: Blue=Index 0 (Forward from saddle), Red=Index 1 (Backward from saddle) }
    \vspace{-2em}
\end{figure}

We \jaledit{also tested a second} analytic dataset which we call Cosine.  The equations for it are from Dey et al.~\cite{DLW07}:
\begin{equation}
\label{eq:cosineVectorField}
    f(x,y) = 50 \change{ \begin{pmatrix}
        \cos(0.06\sqrt{x^2 + y^2}) \\
        \cos(0.001xy)
    \end{pmatrix}
    }
\end{equation}
This field is defined over a domain for $x,y \in [-120,120]$ and sampled every 1 unit.  \jaledit{The structure of this field creates slowly spiralling critical points near the top and bottom center that are surrounded by the separatrices from the saddles near the left and right middle.}  
Figs.~\ref{fig:cosineOurs} and \ref{fig:cosineFast} shows our algorithm's results compared to the discrete representation produced by FastCVT's algorithm. The results show similarities to each other and the underlying flow. There are only a few extra discrete critical points (produced by both datasets) shown as overlapping spheres in each algorithm.
We also see that highly rotating features can be assigned to critical simplices of different indices between the two algorithms, depending on if the alignment of the mesh and discrete vectors produces a bounding closed orbit.

\jaledit{Finally, to comprehensively evaluate the algorithm's performance across diverse conditions, 
a numerical experiment was conducted. 
We generated 1,000 vector fields on a regular grid ($256\times256$) with components randomly sampled in $[-1,1]^2$.  These fields were then smoothed with Gaussian blending to yield approximately 200 PL critical points each.  Our goal was to see how these critical points manifest in the discrete setting, and in particular how nearby those critical points occur.}

\jaledit{The primary outcomes of these experiments are summarized in Table~\ref{tab:experiment}. Across all datasets, a total of 201,131 PL critical points were 
generated. 
Both our algorithm and FastCVT  were applied to generate discrete critical points, ranging from 255 to 425 points per field. Each PL critical point was then evaluated to determine if a discrete critical point of the appropriate type (saddles mapped to critical simplices of dimension 1, non-saddles to critical simplices of dimension 0 or 2) was within varying proximity levels.}

\change{The proximity levels were defined as follows: the first level included any simplex within the triangle containing the PL critical point; the second level comprised simplices containing a vertex of the PL triangle (i.e., within the star of one of the triangle's vertices); the final two levels considered simplices further away, using Euclidean distance from the PL critical point to the barycenter of the nearest discrete critical simplex($c(\sigma)$), either within 2 or 3 units away\jaledit{, which includes the simplices in the 2 and 3 neighborhoods}.}

\jaledit{The results indicate that both algorithms generally are quite similar.  While our algorithm captures slightly more than FastCVT, both find 98\% of the critical points within the star of the triangle containing the PL critical point.
Examination of PL critical points that lacked a nearby discrete representation  typically revealed some kind of degeneracy, including relatively small  determinant of the Jacobian matrix, eigenvectors with shallow angles (i.e., nearly linearly dependent), or their position in close proximity to the boundary of their triangle. However, unlike with discrete critical points manifesting near PL critical points in scalar fields, one cannot make a universal observation.}

\begin{table}[!t]
    \centering
    \caption{
    \change{Results of experiments conducted on 1,000 random vector fields, having a total of 201,131 PL critical points. The first column shows the total number of discrete critical points generated by both algorithms.  Remaining columns count how many PL critical points have a corresponding discrete critical point.  ``In $\Delta$'' looks within the triangle $\Delta$ containing the PL critical point (the triangle, its edges, and its vertices).  ``In $\St(\Delta)$'' looks at the simplices in the stars the triangle's vertices.  ``Under $d$'' assesses whether the barycenter $c(\sigma)$ of a discrete critical simplex is within a Euclidean distance of $d$ units from the PL critical point.
    } 
    \label{tab:experiment}
    }
    \footnotesize
    \begin{tabular}{|l||r|r|r|r|r|}
        \hline
        Algorithm & \# Disc.~CP & In $\Delta$ & In $\St(\Delta)$  & Under 2 & Under 3 \\
        \hline
        \hline
        OSA     & 330,518 & 146,111 & 197,725 & 199,889 & 200,105 \\
        \hline
        FastCVT & 341,346 & 141,338 & 197,074 & 199,402 & 199,790 \\
        \hline
    \end{tabular}
\end{table}

\begin{figure*}[tb] 
   \centering
   \begin{subfigure}[t]{0.49\linewidth}
   \includegraphics[alt={A map of the oceans with two regions zoomed in with sharp curves in flow.},width=\linewidth]{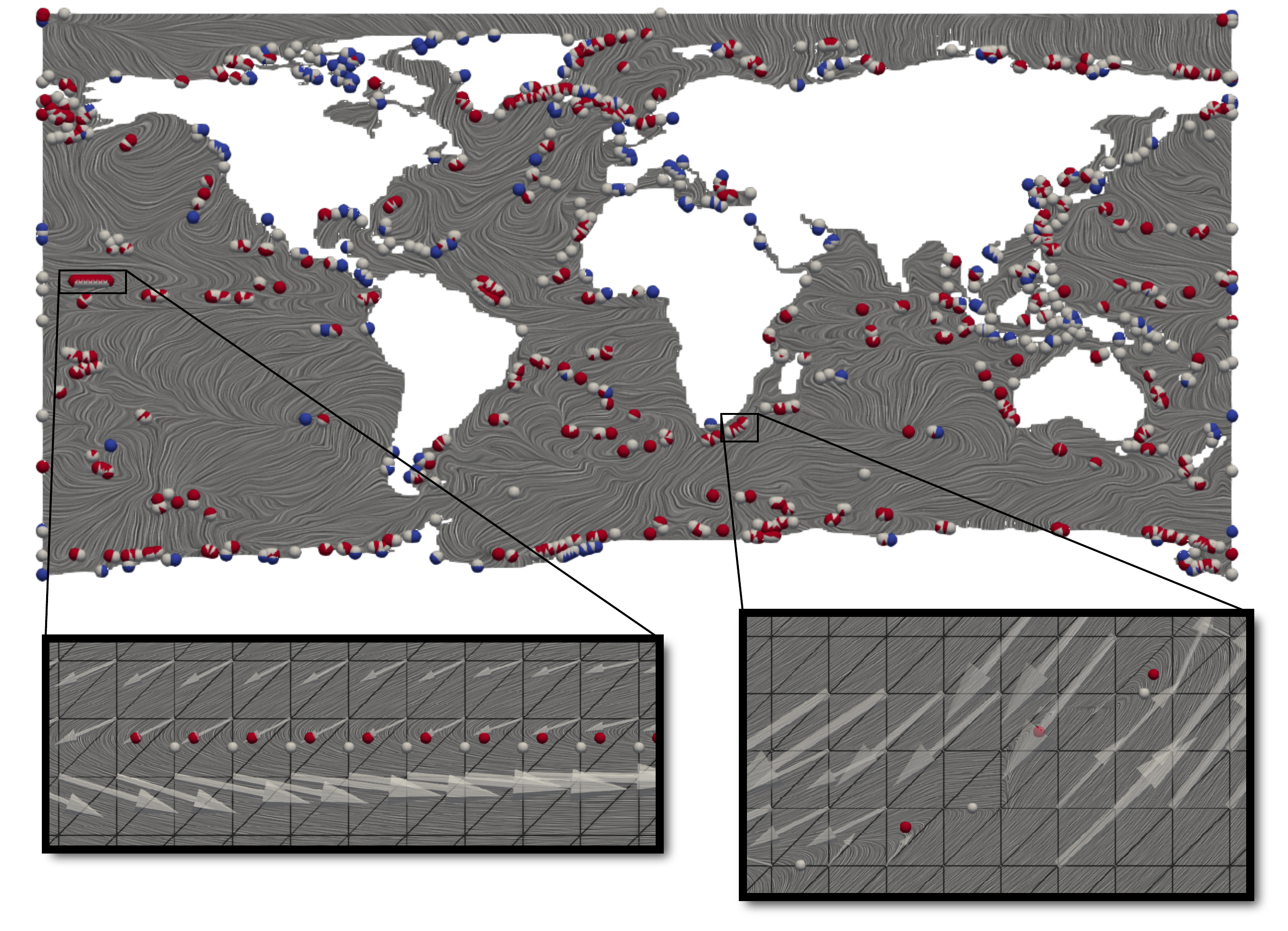}
   \vspace{-0.5ex}
    \caption{\label{fig:chainingOurs} OSA}
    \end{subfigure}
    \begin{subfigure}[t]{0.49\linewidth}
   \includegraphics[alt={A map of the oceans with two diferent regions zoomed in with sharp curves in the flow.},width=\linewidth]{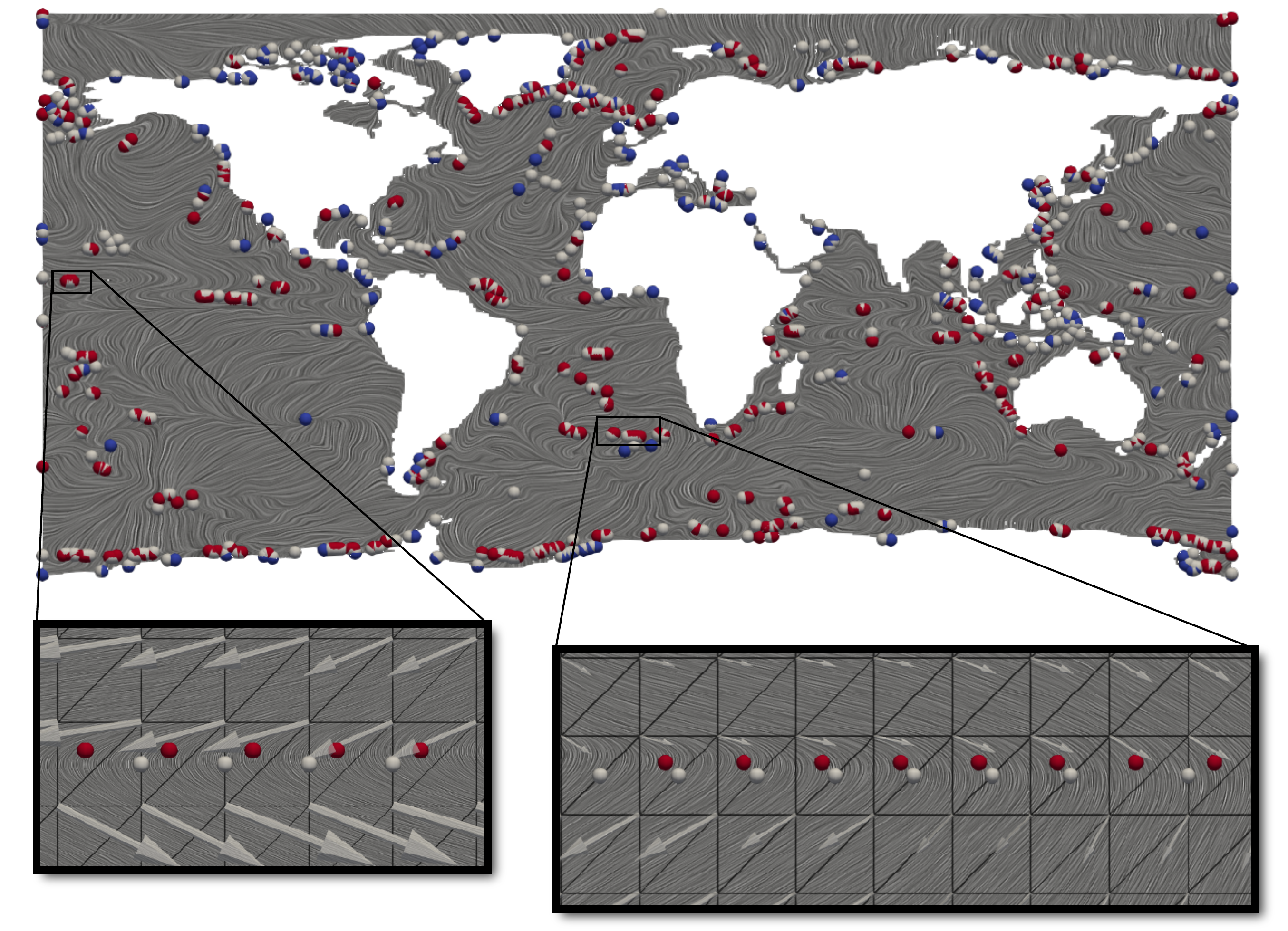}
   \vspace{-0.5ex}
    \caption{\label{fig:chainingFast} FastCVT}
    \end{subfigure}
    \vspace{-2ex}
    \caption{Extraction of critical points from Ocean dataset using (a) our algorithm and (b) FastCVT. The zoom-ins demonstrate the ``chaining'' effect produced by both algorithms, nearby sharp changes in direction where the mesh does not have sufficient resolution to represent the flow. }
\end{figure*}

\change{Looking at the extra discrete critical simplices, \jaledit{many of them were on the boundary.}}  \jaledit{In addition, we found} an interesting artifact of discrete algorithms related to how the mesh discretization interacts with the underlying flow. 
We refer to this as \emph{chaining}, and observe it occurs when sharp curves in the flow are unable to be represented by the size of the provided mesh in a suitable manner. 
Due to the rapid change in direction, a sequence of index $p$ and $p+1$ critical points are generated, usually around a sharp curve in the flow. 
In complex flows such as the Ocean dataset, this characteristic occurs in multiple regions. 
Interestingly this characteristic is also a property of the extraction performed by FastCVT. 
Fig.~\ref{fig:chainingOurs} shows the extraction of critical points by our algorithm with a zoomed-in image displaying the chaining. 
Similarly, Fig.~\ref{fig:chainingFast} shows the extraction using FastCVT and a zoomed-in image to also display chaining. 
FastCVT observed that these critical points without associated piecewise linear 0's are low weight in their algorithm and can be cancelled quickly. 
These examples motivate the necessity of simplification, both to remove small scale features in the data as well as to account for discretization artifacts when switching from piecewise linear to discrete representations.

\subsection{Simplification Analysis}

\begin{figure}[!t]
   \centering
   \begin{subfigure}[c]{0.33\linewidth}
      \includegraphics[alt={Flow with added noise to produce more spheres.},width=\linewidth]{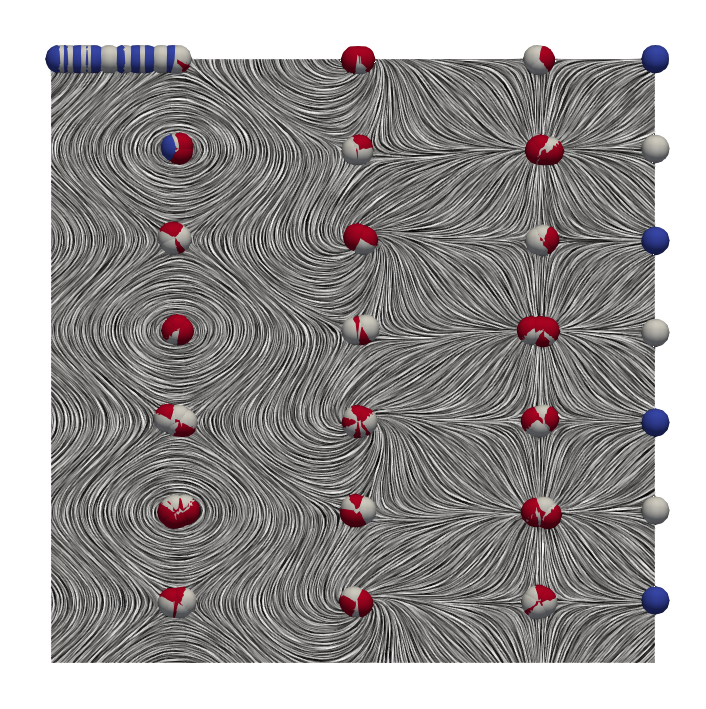}
      \vspace{-1em}
      \caption{\label{fig:simpleUnsimplified}Unsimplified}
      \vspace{-1em}
   \end{subfigure}
   \hfill
   \begin{subfigure}[c]{0.33\linewidth}
      \includegraphics[alt={Same flow with only 27 spheres.},width=\linewidth]{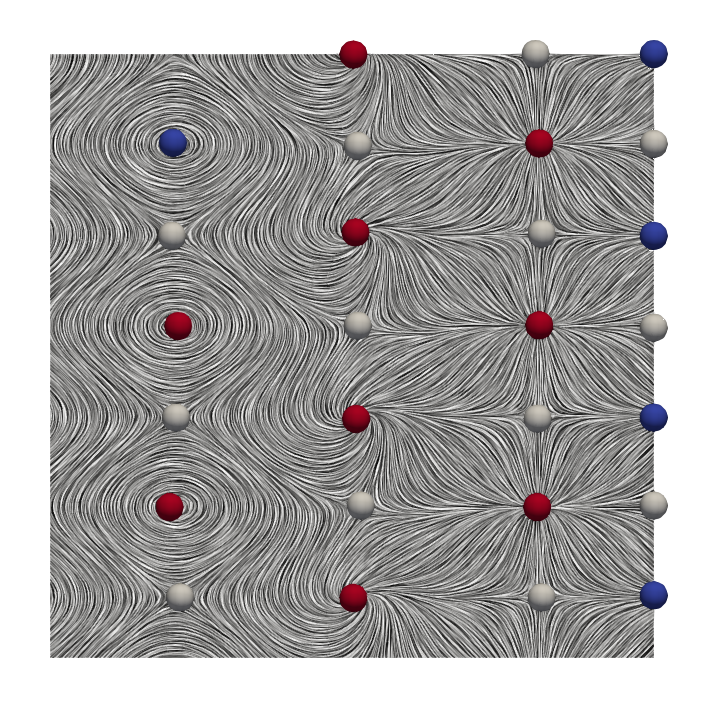}
      \vspace{-1em}
      \caption{\label{fig:simpleNoiseSimplified}Simplified}
      \vspace{-1em}
   \end{subfigure}
   \hfill
   \begin{subfigure}[c]{0.32\linewidth}
      \includegraphics[alt={Line chart plotting cancellation cost at different number of critical points.},width=\linewidth]{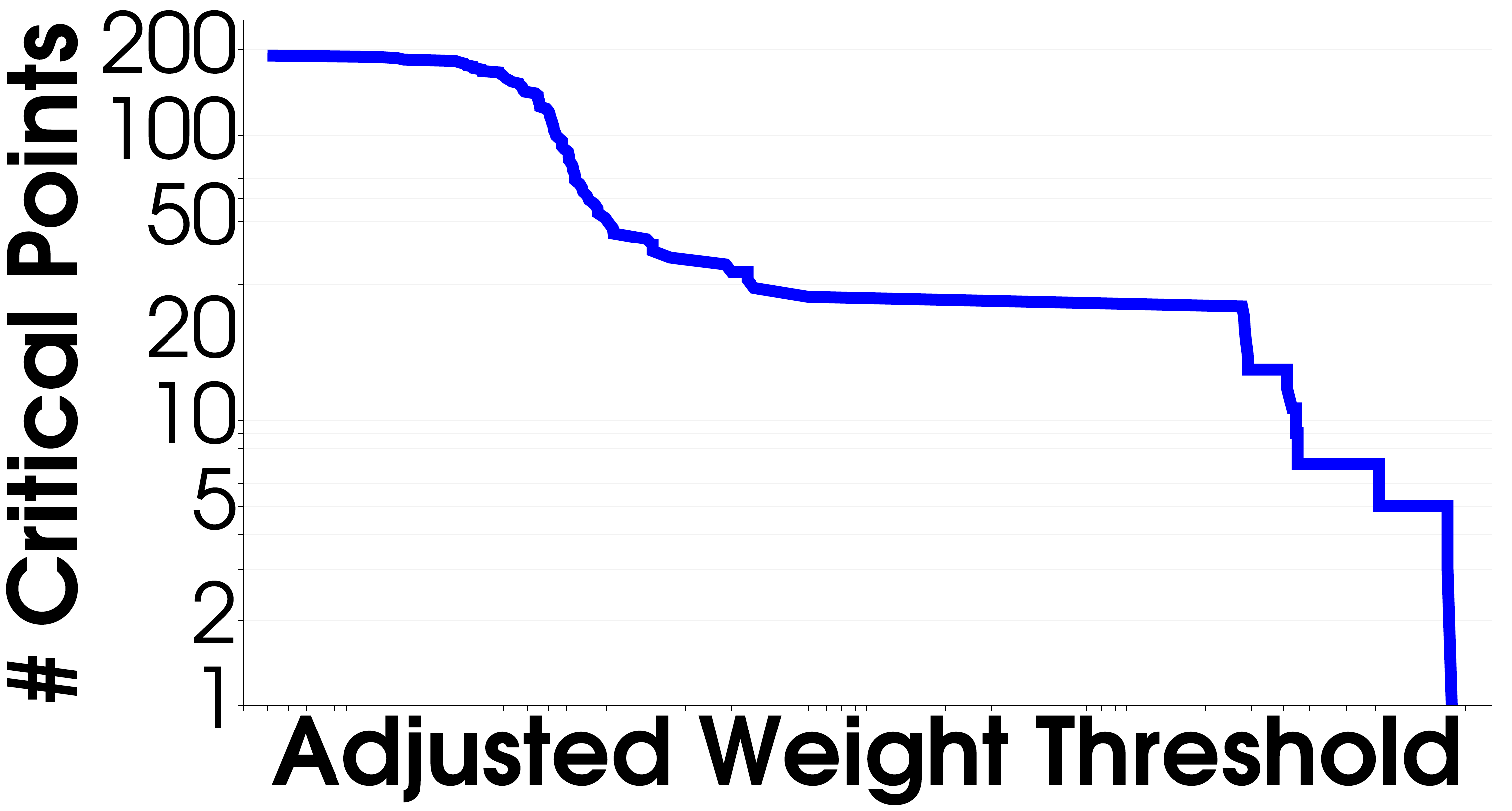}
      \caption{\label{fig:simpleNoiseCurve}Weight Curve}
   \end{subfigure}
    \caption{ Evaluation of vector field produced by Equation \ref{eq:vectorField} and added noise \change{$[-0.3,0.3]^2$}. (a) Our discrete critical points produced by the input. By identifying the flat region in the weight curve (c), we apply simplification (b) using Algorithm \ref{alg:simplify} to the select the 27 critical points associated with the significant flat region, removing the noise.
    }
\end{figure}

\change{To show our simplification method, }we compared 
\change{a} dataset for using simplification guided by weight of simplified pairs. 
\change{W}e revisit the \change{Changes} dataset, adding to each vector a random $\epsilon \in \change{[-0.3, 0.3]^2}$ to produce noise.
The choice of noise was enough to produce noisy critical points as shown in Fig.~\ref{fig:simpleUnsimplified}. The simplification method is then applied to a target \change{27} critical points which corresponds to the large flat region of line chart in Fig.~\ref{fig:simpleNoiseCurve}, and the result is shown in Fig.~\ref{fig:simpleNoiseSimplified}. 
The simplification \change{resulted in} the large features \change{being} preserved although not necessarily in the same location as the original field. 

\begin{figure}[!t] 
\centering
\includegraphics[alt={Complex flow with the center zoomed in using different thresholds based on the weight curve displayed.},width=0.98\linewidth]{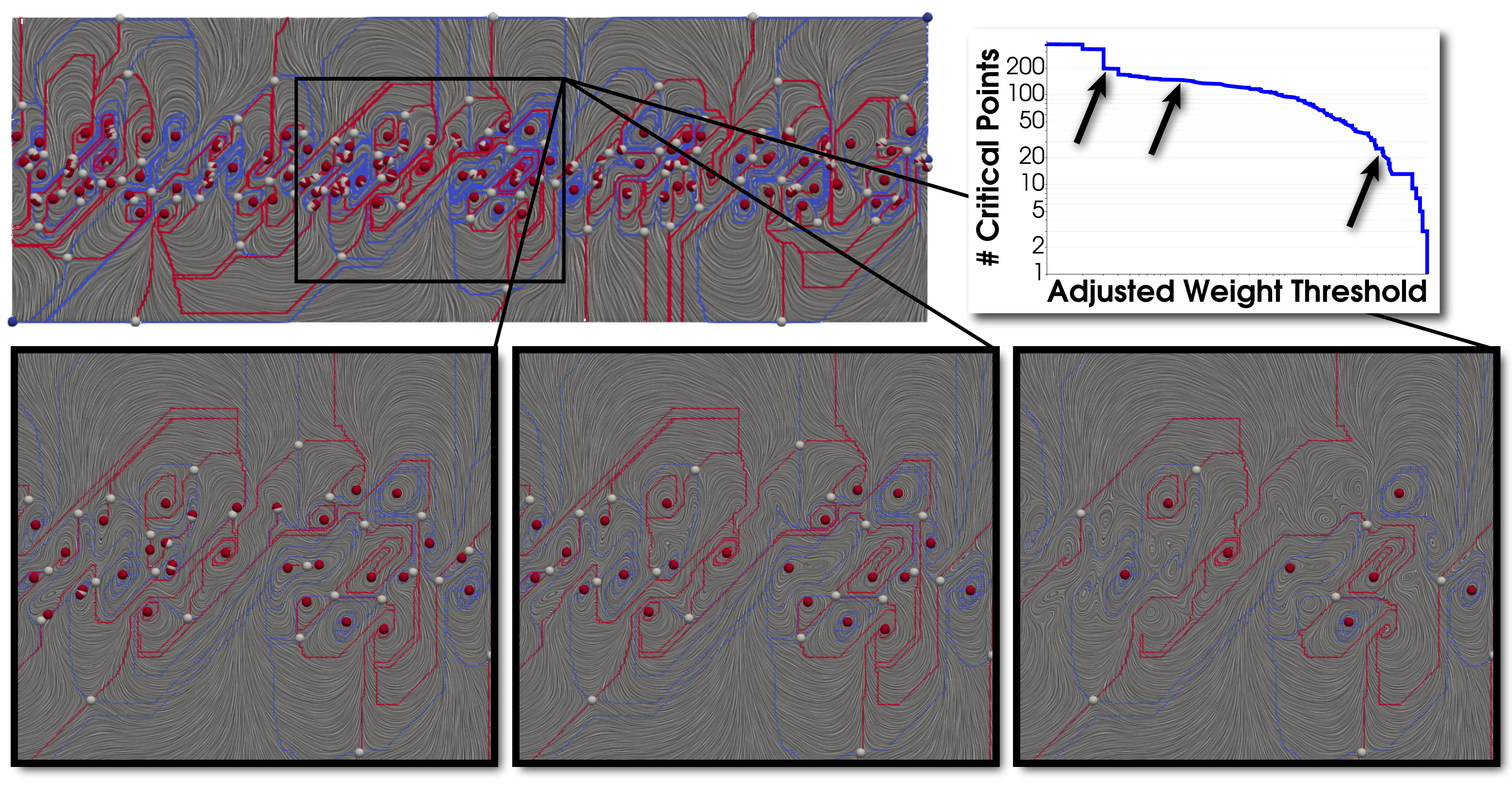} 
   \vspace{-0.5ex}
    \caption{\label{fig:RTTotal} We simplify the RT70 dataset to three different levels (bottom row) guided by the Weight curve at the three indicated thresholds.}
    \vspace{-3ex}
\end{figure}

Our simplification algorithm does not just simplify noisy features in the dataset, but it can also simplify the discrete vector field to eliminate the small features of different weights.  
We simplify the RT70 dataset to three different weight thresholds, shown in Fig.~\ref{fig:RTTotal}. The flatter parts of the weight curve identify more stable regions to simplify to. 
The figure shows the original extraction result of running our algorithm on the RT70 dataset, and then the zoom-ins show three different simplification levels of a particular region in the center.
We simplify to thresholds annotated by arrows of corresponding position (left, center, right).  Simplification has also been applied to the example shown in Fig.~\ref{fig:teaser}, setting a threshold of 2,000 critical points to view the largest features in the data as many of those discrete critical points were generated as boundary artifacts. 

\section{Discussion\label{sec:discussion}}

Our approach of local evaluation results in extremely promising compute times, while producing results that are comparable to the previously existing FastCVT\cite{fastCVT}.
Even though FastCVT is a parallelizable approach, we report times as executed on a single CPU for both our work and FastCVT, so as to be compared on equal platforms.
That said, our discrete construction is embarrassingly parallel as each outward star can be processed independently.
Meanwhile, our saddle simplification routine would require a bit more work to parallelize, as the order in which separatrices are cancelled has an impact.

While we focus on two-dimensional \jaledit{triangulated} vector fields in this work, the algorithm is built on Robins et al.~\cite{LowerStar}, which would generally apply to a simplicial complex of any dimensions, \jaledit{including} 
two-dimensional manifolds embedded in three dimensions as well as three-dimensional domains. 
For two-dimensional manifolds imbued with three-dimensional vectors, we may need to reconsider Eq.~\ref{eq:edgeFlow}.
Most of the implementation choices we made should apply directly in three dimensions, including our definition of outward stars.
\jaledit{Note that} 
our approach for saddle based cancellation would struggle in higher dimensions, as saddle-saddle pairs would be difficult to extract, rank, and remove.
Nevertheless, we feel this direction has promise, and we might be able to develop additional notions that use the same definition of weight to get closer to an equivalent of persistence simplification for volumetric fields. 
\jaledit{Furthermore, we expect the algorithm can also generalize to an arbitrary cell complex (Robins et al.'s algorithm is not restricted to simplices), although our outward star definition (\ref{eq:outStar}) may require additional adaptation.}

Interestingly, for scalar fields, homotopy expansion leads to a key connection between discrete and piecewise linear fields: one can guarantee that all piecewise linear critical points (which for scalar fields must be on vertices) will have a discrete critical simplex in their star~\cite{LowerStar}.
As piecewise linear critical points for a vector field typically fall on the interior of a simplex, we can make no such guarantee. 
In practice though, we observe that most piecewise linear critical points manifest as a nearby discrete critical simplex.
Nevertheless, homotopy expansion does appear to help reduce spurious critical points.
Like with FastCVT, our approach can produce extraneous critical points that are the result of rapid changes in direction in the vector field, which for us manifest as chains that typically can be removed with a small amount of saddle cancellation.

\change{When our algorithm is applied to a gradient field derived from a scalar field, we cannot guarantee the resulting discrete vector field will accurately represent discrete critical points nearby the scalar critical points or be a true gradient field (i.e., loop-free). This behavior is due to the computed gradient vector being merely a derived approximation of the uphill direction. However, we observed that the algorithm performs more accurately, resembling a discrete gradient calculation, when the critical points of the scalar field are well-separated. }

Our algorithm has a couple of components that we view as modular, and further exploration might help to improve the resulting discrete flow.
\jaledit{How we order elements in the priority queues will have a more dramatic impact in three dimensions.}
While our sort order reflects a close match to the discrete gradient computation for a scalar field, it is possible that one could instead dynamically compute a value for a simplex, based on dimension and option value, before it is inserted into the priority queues, where option value is based on previous choices made by the algorithm.
Finally, we observed that choosing to break ties when evaluating Eq.~\ref{eq:edgeFlow} had an overall limited impact.  
That said, other ideas for edge weight formulation include a non-local traversal of the flow which is not guaranteed to terminate, averaged vectors in the surrounding region of that edge, or additional case handling when the orthogonal component of the averaged vector is larger than the resulting edge weight.

\acknowledgments{%
\vspace{-0.5ex}
\footnotesize
This work is partially supported 
by the U.S. Department of Energy, Office of Science,
under Award Number(s)
DE-SC-0019039 and
the European Commission grant
ERC-2019-COG \emph{``TORI''} (ref. 863464, \url{https://erc-tori.github.io/}).
}

\bibliographystyle{abbrv-doi-hyperref}

\bibliography{paper}



\end{document}